\documentclass[12pt]{article}
\usepackage{amsmath, amssymb,amsthm}
\usepackage{psfrag,epsf}
\usepackage{enumerate}
\usepackage{natbib}
\usepackage{url} % not crucial - just used below for the URL 

\usepackage{graphicx, float}
\usepackage[flushleft]{threeparttable}
\usepackage{multirow}
\usepackage[english]{babel}
\usepackage{tabulary}
\usepackage{subcaption}
\usepackage[gen]{eurosym}
\usepackage{accents}
\usepackage{graphicx}
\usepackage{booktabs}
\usepackage[gen]{eurosym}

\DeclareMathOperator*{\argmin}{arg\,min}

\newcommand{\blind}{1}

\newtheorem{remark}{Remark}
\newtheorem{lemma}{Lemma}
\newtheorem{theorem}{Theorem}

\begin{document}

\def\spacingset#1{\renewcommand{\baselinestretch}%
{#1}\small\normalsize} \spacingset{1}

%%%%%%%%%%%%%%%%%%%%%%%%%%%%%%%%%%%%%%%%%%%%%%%%%%%%%%%%%%%%%%%%%%%%%%%%%%%%%%

\if1\blind
{
  \title{\bf Overfitting Reduction in Convex Regression}
  \author{Zhiqiang Liao\\
    Aalto University, Espoo, Finland \\
    Sheng Dai\thanks{Address for correspondence: Sheng Dai, Turku School of Economics, Rehtorinpellonkatu 3, 20500 Turku, Finland. Email: \texttt{sheng.dai@utu.fi}.} \\
    University of Turku, Turku, Finland \\
    Eunji Lim \\
    Adelphi University, New York, USA \\
    Timo Kuosmanen \\
   University of Turku, Turku, Finland \\
    }
  \maketitle
} \fi

\if0\blind
{
  \bigskip
  \bigskip
  \bigskip
  \begin{center}
    {\LARGE\bf Overfitting Reduction in Convex Regression}
\end{center}
  \medskip
} \fi

\bigskip
\begin{abstract}
\noindent Convex regression is a method for estimating the convex function from a data set. This method has played an important role in operations research, economics, machine learning, and many other areas. However, it has been empirically observed that convex regression produces inconsistent estimates of convex functions and extremely large subgradients near the boundary as the sample size increases. In this paper, we provide theoretical evidence of this overfitting behavior. To eliminate this behavior, we propose two new estimators by placing a bound on the subgradients of the convex function. We further show that our proposed estimators can reduce overfitting by proving that they converge to the underlying true convex function and that their subgradients converge to the gradient of the underlying function, both uniformly over the domain with probability one as the sample size is increasing to infinity. An application to Finnish electricity distribution firms confirms the superior performance of the proposed methods in predictive power over the existing methods.
\end{abstract}

\noindent%
{\it Keywords:}  Machine learning, Convex regression, Overfitting, Lipschitz regularization, Weight restriction
\vfill

\newpage
\spacingset{1.8} % DON'T change the spacing!
% Text of your paper here

\section{Introduction}

Convex regression (CR), a classical nonparametric regression method with shape constraints dating back to \cite{hildreth1954point}, has attracted growing interest in operations research \citep{lee2013a, balazs2015near}, econometrics \citep{kuosmanen2008representation, yagi2020shape}, statistical learning \citep{blanchet2019multivariate, bertsimas2021sparse}, and many other areas. One of the great advantages of CR over other nonparametric techniques is that it does not require any tuning parameters (e.g., the bandwidth parameter as in kernel estimation). Consequently, many applications of CR can be found in various fields such as decision analysis \citep{lim2012consistency}, portfolio selection \citep{hannah2013multivariate}, productivity analysis \citep{kuosmanen2012stochastic2}, energy regulation \citep{kuosmanen2020conditional}, and environmental policy \citep{kuosmanen2021shadow}. 

Even though the CR estimator is a natural candidate for an estimator of convex functions, it produces inconsistent estimates of the underlying true convex function and extremely large subgradients near the boundary \citep{ghosal2017univariate}. This undesirable overfitting problem has been observed by, e.g., \cite{lim2012consistency}, \cite{mazumder2019a}, and \cite{liao2024convex}. Recently, an empirical application of CR to the estimation of the cost frontier function for Finnish energy distribution firms yielded extremely large subgradients near the boundary \citep{kuosmanen2022}, which deteriorated the performance of CR in evaluating their incentive regulatory model for electricity firms. Although these studies have shown that the overfitting problem tends to occur in CR, they do not provide any theoretical evidence that explains the existence of overfitting. This motivated us to theoretically investigate the overfitting problem in the framework of CR. 

Several attempts have been made to avoid overfitting in the context of CR. For instance, \cite{bertsimas2021sparse} proposed penalized convex regression (PCR) by adding a penalty term to the objective function of the optimization problem. This formulation allows us to limit the overall behavior of CR, but we cannot directly control the magnitude of each subgradient. \cite{mazumder2019a} proposed using Lipschitz convex regression (LCR) by adding additional constraints to the subgradient at each data point and, thus, restricting overfitting in a straightforward sense. However, little is known about how LCR behaves on the boundary of the domain and whether it does not exhibit overfitting behavior near the boundary. Indeed, whether placing a hard bound on the subgradients eliminates the overfitting problem remains undetermined. We answer this question affirmatively by proving that the hard bound on the norm of the subgradients generates an estimator that converges to the underlying function uniformly over the domain, including the boundary, with probability one as the sample size increases to infinity. Furthermore, this estimator has subgradients that converge to the gradient of the underlying function uniformly, including the boundary, with probability one as the sample size increases to infinity; see Theorems \ref{theorem: alcr consistency} and \ref{theorem: wrcr consistency} of this paper. 

We further propose two practical guides on how to reduce overfitting by placing a bound on the subgradients. We first proposed an augmented extension of LCR, which originated from the work of \cite{mazumder2019a}. In the first method, we noticed that we could find a good ``reference'' for the estimator. For example, one can imagine that the linear regression estimator could be a good starting point without exhibiting any overfitting behavior when estimating a convex function. Alternatively, the percentiles of the subgradients of the convex regression estimator can be another good starting point for the search. The second remedy for the overfitting problem is to place the lower and upper bounds on the subgradients, which are referred to as weight restrictions in the literature (see, e.g., \citealp{podinovski2016optimal}). The easiest way to obtain the hard bounds is to get input from decision-makers. The decision-makers are often able to specify the upper and lower bounds of the subgradients. For example, in the application context where a decision-maker is trying to estimate a cost function, the subgradients are interpreted as the marginal costs, and the decision-maker may have a good understanding of how high the marginal cost can be \citep{kuosmanen2021shadow}. Thus, the prespecified values of the upper and lower bounds on the subgradients can be readily obtained from the decision-maker. If such information is unavailable, one can rely on the original subgradients obtained from the CR estimator and use their percentiles as the bounds. Adding an explicit bound helps decision-makers understand how the model works, so they trust the model when they use it. 

Our main contributions can be summarized as follows.

\textit{Theoretical evidence of the overfitting behavior}: We study the theoretical properties of the CR estimator near the boundary. Although the consistency of the CR estimator in the interior of the domain has been proven by \cite{seijo2011nonparametric} and \cite{lim2012consistency}, its convergence near the boundary of the domain is not guaranteed. Furthermore, the subgradient of the CR estimator is observed to be very large at the boundary \citep{mazumder2019a}. \cite{ghosal2017univariate} have shown that the subgradient of the CR estimator is unbounded in probability at the boundary in the univariate case, but no such result has been obtained in the multivariate case. In this paper, we establish this result in the multivariate setting and prove the unboundedness of the subgradient of the CR estimator at the boundary of its domain. To the best of our knowledge, this is the first attempt to systematically study the overfitting behavior of CR under the multivariate setting.

\textit{Our proposed estimators}: We propose two new estimators, the augmented Lipschitz convex regression (ALCR) and weight-restricted convex regression (WRCR) estimators. Each of them provides a simple and practical way to bound the subgradients of the fitted convex function and, therefore, can alleviate overfitting. We further prove that these estimators and their subgradients are strongly consistent uniformly over the domain. Numerical results in Sections \ref{sec: MC} and \ref{sec: application} display superior performance of the ALCR and WRCR estimators compared to some existing estimators. While other techniques for avoiding overfitting exist, our proposed estimators provide both theoretical and practical evidence supporting the performance of the proposed hard-bounded approaches.

\textit{Application}: In Finland, the Energy Authority (EV) has systematically applied CR to implement economic incentives for Finnish electricity distribution firms since 2012 \citep{kuosmanen2012stochastic,kuosmanen2020conditional}. However, the conventional CR models resulted in high estimated marginal costs (or subgradients) for several big firms; see, e.g., the descriptive statistics in Table 2 of \cite{kuosmanen2012stochastic}. Moreover, the incentives for these firms in future periods always rely on the predictive performance of the CR model based on historical data. That is, the effects of the incentive program are subject to the predictive power of the CR model. The empirical study shows that we can address the overfitting problem using the proposed estimators and create the right incentives for regulated firms. 

The rest of this paper is organized as follows. In Section \ref{sec: pre}, We introduce the problems of convex regression and overfitting reduction, along with relevant notation and definitions. In Section \ref{sec: main}, we provide some theoretical evidence on the overfitting behavior of the CR estimator. In Section \ref{sec: alcr}, we introduce the proposed ALCR estimator and investigate its statistical properties. The WRCR estimator and its properties are developed and analyzed in Section \ref{sec: wrcr}. In Section \ref{sec: MC}, we perform Monte Carlo studies to compare the performance of the proposed estimators to that of some existing estimators. An empirical application of our proposed methods to Finnish electricity distribution firms is presented in Section \ref{sec: application}. In Section \ref{sec: conclusion}, we conclude this paper with suggestions on future research avenues.

\section{Preliminaries}\label{sec: pre}
\subsection{Convex regression}

CR can be formally described as follows. Given data $(\boldsymbol{x}_1,y_1), \cdots, (\boldsymbol{x}_n, y_n)$, we assume
\begin{equation} \label{eq: non}
    y_i = f_0(\boldsymbol{x}_i) + \varepsilon_i 
\end{equation}
for $i = 1, \cdots, n$, where $\boldsymbol{x}_i \in \Omega \subset \mathbb{R}^d$, $f_0:\Omega \rightarrow \mathbb{R}$ is the underlying function to be estimated, and the $\varepsilon_i$'s are the error terms satisfying $\mathbb{E}[\varepsilon_i]=0$ and $\text{Var}[\varepsilon_i]=\sigma^2 < \infty$ for $i = 1, \cdots, n$. Even though $f_0$ cannot be observed exactly, it is known to be convex. In the context of CR, our goal is to estimate $f_0$ by fitting a convex function to the given $n$ observations. In particular, we estimate $f_0$ by minimizing the sum of squared errors:
\begin{equation} \label{lse}
    \frac{1}{n}\sum_{i=1}^n (y_i - f(\boldsymbol{x}_i))^2
\end{equation}
over $f \in \mathcal{F}$, where $\mathcal{F} \triangleq \{f:\Omega\rightarrow \mathbb{R} \mbox{ such that } f \mbox{ is convex}\}$ is the class of all real-valued convex functions over $\Omega$. 

In (\ref{lse}), we try to fit a convex function $f$ to the data points $(\boldsymbol{x}_1, y_1), \cdots, (\boldsymbol{x}_n, y_n)$ and find the one with the least squares, so (\ref{lse}) is an optimization problem over functions and it appears to be infinite-dimensional. However, the infinite-dimensional problem (\ref{lse}) can be reduced to a finite-dimensional quadratic program where the decision variables are the function values and subgradients of the fitted convex function $f$ at the $\boldsymbol{x}_i$'s. The following quadratic program, in the decision variable $f_1, \cdots, f_n \in \mathbb{R}$ and $\boldsymbol{\beta}_1, \cdots, \boldsymbol{\beta}_n \in \mathbb{R}^d$, is one such formulation:

\begin{equation}\label{cr_new_formulation}
\begin{aligned}
    \min \quad & \frac{1}{n}\sum_{i=1}^{n} (y_i - f_i)^2  &{}&  \\
    \mbox{s.t.}\quad &f_j \geq f_i + \boldsymbol{\beta}_i^T (\boldsymbol{x}_j - \boldsymbol{x}_i) &\quad& i, j = 1, \cdots, n;
\end{aligned}
\end{equation}
see, e.g., \cite{seijo2011nonparametric} and \cite{lim2012consistency}. In (\ref{cr_new_formulation}), $f_i$ and $\boldsymbol{\beta}_i$ represent the value of the fitted function $f$ and a subgradient of $f$, respectively, for $i= 1, \cdots, n$. The set of constraints in (\ref{cr_new_formulation}), namely shape constraints, enforces the fitted function $f$ to be convex.

In the cases of production/cost function estimation, an alternative formulation is commonly used in the literature as the returns to scale need to be specified; see, e.g., \cite{afriat1972efficiency}, \cite{banker1993maximum}, and \cite{kuosmanen2010data}. Formally, the following quadratic program, in the decision variables  $\alpha_1, \cdots, \alpha_n \in \mathbb{R}$, and $\boldsymbol{\beta}_1, \cdots, \boldsymbol{\beta}_n$ $\in \mathbb{R}^d$, can be used:
\begin{equation}\label{cr}
\begin{aligned}
    \min \quad & \frac{1}{n}\sum_{i=1}^{n} \varepsilon_i^2  &{}&  \\
    %\mbox{s.t.}\quad & y_i = \alpha_i + \boldsymbol{\beta}_i^T \boldsymbol{x}_i + \varepsilon_i &\quad& i = 1, \cdots, n \\
    \mbox{s.t.}\quad &y_i = \alpha_i + \boldsymbol{\beta}_i^T \boldsymbol{x}_i + \varepsilon_i  &\quad& i = 1, \cdots, n \\
   \quad &  \alpha_i + \boldsymbol{\beta}_i^T \boldsymbol{x}_i \geq \alpha_j + \boldsymbol{\beta}_j^T \boldsymbol{x}_i &{}&  i, j = 1, \cdots, n;
\end{aligned}
\end{equation}
see \cite{kuosmanen2008representation} for details. In (\ref{cr}), $\alpha_i$ is a constant representing the intercept and $\boldsymbol{\beta}_i$ denotes the slope coefficients (or subgradient of $f$), at $\boldsymbol{x}_i$ for $i= 1, \cdots, n$. By solving (\ref{cr_new_formulation}) or (\ref{cr}), we can numerically compute a solution to (\ref{lse}), which is referred to as the CR estimator. For simplicity of notation, we will use the formulation in \eqref{cr_new_formulation} in the following text, and we will consider problem \eqref{cr} in Section \ref{sec: application} for cost function estimation. 

\subsection{Overfitting reduction}
Overfitting is a common issue in machine learning where the fitted models are overly complex and fail to capture the underlying true patterns \citep{lever2016points}. This leads to a model that performs perfectly on training data but poorly on unseen data, resulting in poor out-of-sample performance. For example, high-order polynomial regression is more flexible than its low-order counterpart and, hence, is more prone to suffer from the overfitting problem. 

To avoid overfitting in CR, we often need to restrict the model flexibility so that the estimator does not lead to an extreme model with large subgradients, especially near the boundary. A widely used way to do so is to add the penalty (regularization) term to the optimization problem \eqref{cr_new_formulation}. For example, \cite{bertsimas2021sparse} proposed PCR by adding $L_2$-norm regularization (see also in \citealp{lim2021consistency}). That is, we minimize the sum of squared errors and squared norms of the subgradients as follows:
\begin{equation} \label{lse_pcr}
    \min \frac{1}{n}\sum_{i=1}^n (y_i - f(\boldsymbol{x}_i))^2 + \frac{\lambda_n}{n}\sum_{i = 1}^n \|\boldsymbol{\beta}_i\|^2
\end{equation}
over $f \in \mathcal{F}$ for a given sequence of nonnegative real numbers $(\lambda_n: n \geq 1)$. The infinite-dimensional problem (\ref{lse_pcr}) can be reformulated as the following finite-dimensional quadratic program in the decision variables $f_1, \cdots, f_n \in \mathbb{R}$ and $\boldsymbol{\beta}_1, \cdots, \boldsymbol{\beta}_n \in \mathbb{R}^d$:
\begin{equation}\label{pcr}
\begin{aligned}
    \min \quad & \frac{1}{n}\sum_{i=1}^{n} (y_i - f_i)^2 + \frac{\lambda_n}{n}\sum_{i = 1}^n \|\boldsymbol{\beta}_i\|^2  &{}&  \\
    \mbox{s.t.}\quad &f_j \geq f_i + \boldsymbol{\beta}_i^T (\boldsymbol{x}_j - \boldsymbol{x}_i)  &\quad& i, j  = 1, \cdots, n.
\end{aligned}
\end{equation}
The solution $\hat{f}_{1, p}, \cdots, \hat{f}_{n, p}, \hat{\boldsymbol{\beta}}_{1, p}, \cdots, \hat{\boldsymbol{\beta}}_{n, p}$ to (\ref{pcr}) exists uniquely \citep
{lim2021consistency}. We define $\hat{f}_{n, p}:\Omega \rightarrow \mathbb{R}$  by
\begin{equation}
\label{construction_of_PCV_estimator}
    \hat{f}_{n, p}(\boldsymbol{x}) = \max_{1 \leq i \leq n}\{\hat{f}_{i, p} + \hat{\boldsymbol{\beta}}_{i, p}^T(\boldsymbol{x} - \boldsymbol{x}_i)\}
\end{equation}
for $\boldsymbol{x} \in \Omega$ and refer to $\hat{f}_{n, p}$ as the PCR estimator.

Instead of imposing $L_2$-norm on all subgradients $\sum_{i = 1}^n \|\boldsymbol{\beta}_i\|^2$, \cite{mazumder2019a} apply the $L_2$-norm Lipschitz regularization on each subgradient to manage the overfitting problem. For a given constant $L>0$, LCR is then formulated as
\begin{equation}\label{lcr}
\begin{aligned}
    \min \quad & \frac{1}{n}\sum_{i=1}^{n} (y_i - f_i)^2 &{}&  \\
    \mbox{s.t.}\quad &f_j \geq f_i + \boldsymbol{\beta}_i^T (\boldsymbol{x}_j - \boldsymbol{x}_i)  &\quad& i, j  = 1, \cdots, n.\\
    &\|\boldsymbol{\beta}_i\| \leq L &{}&  i = 1, \cdots, n.
\end{aligned}
\end{equation}
Similarly, we can define a representation function like \eqref{construction_of_PCV_estimator} by replacing the solution $\hat{f}_{1}, \cdots, \hat{f}_{n}, \hat{\boldsymbol{\beta}}_{1}, \cdots, \hat{\boldsymbol{\beta}}_{n}$ with those obtained from problem \eqref{lcr}.

Leveraging the idea of placing hard bounds on the subgradients in \cite{mazumder2019a}, we first consider an extension of the $L_2$-norm Lipschitz regularization in the current paper. That is, we add a reference to the $L_2$-norm such that we aim to regularize the subgradients to the benchmark reference. One can use these ``reference'' values and add the inequality
\begin{equation}
\label{auglip norm}
    \| \boldsymbol{\beta}_i - \boldsymbol{b}_0\| \le L_0
\end{equation}
for $i = 1, \cdots, n$ to the constraints of (\ref{cr_new_formulation}), where $\boldsymbol{b}_0$ represents the reference vector mentioned so far. A key advantage of the new constraints is that they provide a practical way of integrating the prior knowledge of decision-makers (see an empirical study for illustration in Section \ref{sec: application}).

Another practical way to impose a hard bound on the subgradients is adding upper and lower bounds to each subgradient directly as follows:
\begin{equation}\label{weight restriction}
   \boldsymbol{l}_0 \leq \boldsymbol{\beta}_i \leq \boldsymbol{u}_0
\end{equation}
for $n = 1, \cdots, n$, where $\boldsymbol{l}_0 \in \mathbb{R}^d$ and $\boldsymbol{u}_0 \in \mathbb{R}^d$ indicate lower and upper bounds on the subgradient, respectively, and $\leq$ is used componentwise. We will revisit these two hard bounds and investigate their statistical properties when applying to CR estimators in Sections \ref{sec: alcr} and \ref{sec: wrcr}.

\subsection{Notations}
\label{notation}

For $\boldsymbol{x} \in \mathbb{R}^d$, $x(k)$ represents its $k$-th element for $k = 1, \cdots, d$, so $\boldsymbol{x} = (x(1), \cdots, x(k))$. We write $\|\boldsymbol{x}\| = \left\{x(1)^2 + \cdots + x(d)^2\right\}^{1/2}$. The transpose of $\boldsymbol{x} \in \mathbb{R}^d$ is denoted by $\boldsymbol{x}^T$. $\mathbf{0} \in \mathbb{R}^d$ denotes the vector with zeros in all entries. For $\boldsymbol{a}, \boldsymbol{b} \in \mathbb{R}^d$, we write $\boldsymbol{a} \leq \boldsymbol{b}$ if and only if $a(i) \leq b(i)$ for $i = 1, \cdots, d$. We also write $\boldsymbol{a} < \boldsymbol{b}$ if and only if $a(i) < b(i)$ for $i = 1, \cdots, d$.

Let $\Lambda\subset \mathbb{R}^d$ be a convex set. For a convex function $f:\Lambda \rightarrow \mathbb{R}$, we call $\boldsymbol{\beta} \in \mathbb{R}^d$ a subgradient of $f$ at $\boldsymbol{x} \in \Lambda$ if $\boldsymbol{\beta}^T(\boldsymbol{y} - \boldsymbol{x}) \leq f(\boldsymbol{y}) - f(\boldsymbol{x})$ for any $\boldsymbol{y} \in \Lambda$. We denote a subgradient of $f$ at $\boldsymbol{x} \in \Lambda$ by $\mbox{subgrad }f(\boldsymbol{x})$. We call the set of all subgradients at $\boldsymbol{x}$ the subdifferential at $\boldsymbol{x}$, and denote it by $\partial f(\boldsymbol{x})$. For any differentiable function $f:\Lambda \rightarrow \mathbb{R}$, $\nabla f(\boldsymbol{x})$ denotes the derivative of $f$ at $\boldsymbol{x} \in \Lambda$.

\section{The overfitting problem}
\label{sec: main}

In this section, we investigate the behavior of the CR estimator near the boundary of its domain and theoretically explain why it shows the overfitting behavior.

We start by formally defining the CR estimator. In CR, we minimize the sum of squared errors in (\ref{lse}) over $f \in \mathcal{F}$. Problem (\ref{cr_new_formulation}) determines the values of the fitted convex function $f$ at the $\boldsymbol{x}_i$'s (i.e., $f_i$'s) uniquely, but it does not determine the subgradients of the fitted function 
 at the $\boldsymbol{x}_i$'s (i.e., $\boldsymbol{\beta}_i$'s) uniquely; see Lemma 2.5 of \cite{seijo2011nonparametric}. To define the CR estimator uniquely over $\Omega$, we take, among all the $\boldsymbol{\beta}_i$'s solving problem (\ref{cr_new_formulation}), the one with the minimum norm. More precisely, to determine the subgradients of the fitted function at the $\boldsymbol{x}_i$'s uniquely, we solve the following finite-dimensional quadratic program in the decision variables $\boldsymbol{\beta}_1, \cdots, \boldsymbol{\beta}_n \in \mathbb{R}^d$:
\begin{equation}
\label{unique_beta_cr}
\begin{aligned}
    \min \quad & \sum_{i=1}^{n} \|\boldsymbol{\beta}_i\|^2 &{}&  \\
    \mbox{s.t.}\quad &\hat{f}_j \geq \hat{f}_i + \boldsymbol{\beta}_i^T (\boldsymbol{x}_j - \boldsymbol{x}_i) &\quad& i, j = 1, \cdots, n,
\end{aligned}
\end{equation}
where $\hat{f}_1, \cdots, \hat{f}_n$ are the minimizing values of (\ref{cr_new_formulation}). The solution $\tilde{\boldsymbol{\beta}}_1, \cdots,  \tilde{\boldsymbol{\beta}}_n$ to (\ref{unique_beta_cr}) exists uniquely since (\ref{unique_beta_cr}) is a minimization problem with continuous, strictly convex, and coercive objective function over a non-empty and closed set (Proposition 7.3.1 and Theorem 7.3.7 in \citealp{kurdila2006convex}). We now define $\hat{f}_n: \Omega \rightarrow \mathbb{R}$ by
\begin{equation} 
\label{construction_of_CV_estimator}
    \hat{f}_n(\boldsymbol{x}) = \max_{1 \leq i \leq n}\{\hat{f}_i + \tilde{\boldsymbol{\beta}}_i^T(\boldsymbol{x} - \boldsymbol{x}_i)\}
\end{equation}
for $\boldsymbol{x} \in \Omega$ and refer to $\hat{f}_n$ as the CR estimator.

To analyze the properties of the CR estimator near the boundary of $\Omega$, we need the following assumptions:
\begin{enumerate}
\label{assumption:02}
\item[]\textbf{A1.}
$\Omega = [0, 1]^d$ and $\boldsymbol{x}_1, \boldsymbol{x}_2, \cdots$ is a sequence of independent and identically distributed (i.i.d.) $\Omega$-valued random vectors having a common density function $\kappa:\Omega \rightarrow \mathbb{R}$.
\item[]\textbf{A2.}
Given $\boldsymbol{x}_1, \boldsymbol{x}_2, \cdots$, $\varepsilon_1, \varepsilon_2, \cdots$ are i.i.d. random variables with a mean of zero and a non-zero finite variance $\sigma^2$.

\item[]\textbf{A3. (i)}
For any subset $A$ of $\Omega$ with a nonempty interior, $\mathbb{P}(\boldsymbol{x}_1 \in A) > 0$.

\item[]\textbf{\phantom{A3.} (ii)} $\kappa:\Omega \rightarrow \mathbb{R}$ is continuous.

\item[]\textbf{A4.} $\mathbb{E}[f_0 (\boldsymbol{x}_1)^2] < \infty$

\item[]\textbf{A5.}
$f_0:\Omega \rightarrow \mathbb{R}$ is convex.
\item[]\textbf{A6.} $f_0$ is differentiable on $\Omega$.

\item[]\textbf{A7.} There exist $\boldsymbol{b}_0$ and $L_0$ such that $\|\boldsymbol{\beta}- \boldsymbol{b}_0\| \leq L_0$ for any $\boldsymbol{\beta} \in \partial f_0 (\boldsymbol{x})$ and $\boldsymbol{x} \in \Omega$.

\item[]\textbf{A8.}
There exist $\boldsymbol{l}_0, \boldsymbol{u}_0 \in \mathbb{R}^d$ such that 
$\boldsymbol{l}_0 \leq \boldsymbol{\beta} \leq \boldsymbol{u}_0$ for any $\boldsymbol{\beta} \in \partial f_0 (\boldsymbol{x})$ and $\boldsymbol{x} \in \Omega$.
\end{enumerate}

Now, we state the main results of this section. Theorem \ref{theorem: inconsistency} shows that assuming the convexity of $f_0$, the CR estimator is inconsistent in estimating $f_0$ near the boundary of $\Omega$ as $n \rightarrow \infty$. Theorem \ref{theorem: unboundedness} states that assuming the differentiability of $f_0$ over $\Omega$, the subgradients of the CR estimator are unbounded in probability near the boundary of $\Omega$ as $n \rightarrow \infty$. The detailed proofs for Theorems \ref{theorem: inconsistency} and \ref{theorem: unboundedness} are available in the supplementary material to this paper.

\begin{theorem} 
\label{theorem: inconsistency}
Assume \textbf{A1}, \textbf{A2}, \textbf{A3(ii)}, \textbf{A4}--\textbf{A6}. Then there exists $\epsilon_0 > 0$ such that for any $\epsilon \leq \epsilon_0$,
\[\liminf_{n \rightarrow \infty} \mathbb{P} (|\hat{f}_n (\mathbf{0}) - f_0 (\mathbf{0})| > \epsilon) > 0.\]
Thus, $\hat{f}_n(\mathbf{0})$ is inconsistent in estimating $f_0(\mathbf{0})$.
\end{theorem}

\begin{theorem} 
\label{theorem: unboundedness}
Assume \textbf{A1}, \textbf{A2}, \textbf{A3(ii)}, \textbf{A4}--\textbf{A6}.  There exists $\epsilon_1 > 0$ such that for any $M > 0$, we have
\[\liminf_{n \rightarrow \infty} \mathbb{P} \left(\min_{\boldsymbol{\beta} \in \partial \hat{f}_n(\mathbf{0})} \|\boldsymbol{\beta}\| > M\right) > \epsilon_1.\]
Thus, $\min_{\boldsymbol{\beta} \in \partial \hat{f}_n(\mathbf{0})} \|\boldsymbol{\beta}\|$ is not bounded in probability.
\end{theorem}

The unboundedness property in Theorem \ref{theorem: unboundedness} suggests that the estimated subgradients of CR could take very large values near zero, which is one of the boundary points. In the following sections, we propose restricting the estimated subgradients with a known Lipschitz bound (Section \ref{sec: alcr}) and with known upper and lower bounds (Section \ref{sec: wrcr}).

\begin{remark}
\label{remark1}
Theorems \ref{theorem: inconsistency} and \ref{theorem: unboundedness} still hold when $\Omega$ is a hyperrectangle, i.e., $\Omega = [\boldsymbol{a}(1), \boldsymbol{b}(1)]\times \cdots\times[\boldsymbol{a}(d), \boldsymbol{b}(d)] \subset \mathbb{R}^d$ for some $\boldsymbol{a} = (\boldsymbol{a}(1),  \cdots, \boldsymbol{a}(d) \in \mathbb{R}^d$ and $\boldsymbol{b} = (\boldsymbol{b}(1), \cdots, \boldsymbol{b}(d)) \in \mathbb{R}^d$.
\end{remark}

\begin{remark}
\label{remark2}
Lemma 5.1 of \cite{ghosal2017univariate} shows the inconsistency of the CR estimator at the boundary of $\Omega$ in the one-dimensional case when $d = 1$. Thus, Theorems \ref{theorem: inconsistency} and \ref{theorem: unboundedness} can be seen as an extension of their results to the multidimensional case when $d > 1$.
\end{remark}

\section{Augmented Lipschitz convex regression}
\label{sec: alcr}

In this section, we propose the ALCR  estimator, which bounds the norm of the subgradients around a reference vector $\boldsymbol{b}_0$. This method becomes useful when some prior knowledge on the subgradients of $f_0$ is available. We will then show that the ALCR estimator does not show the overfitting behavior by proving its uniform consistency over the entire domain $\Omega$.

Given $\boldsymbol{b}_0 \ge \mathbf{0}$ and $L_0 > 0$, we consider a class of convex functions whose subgradients are uniformly bounded by $L_0$ around $\boldsymbol{b}_0$: 
\begin{equation}\label{class: aug}
    \mathcal{F}_A:= \bigg\{f: \Omega \rightarrow \mathbb{R} \:|\: f  \mbox{ is convex and }\| \boldsymbol{\beta}_i - \boldsymbol{b}_0\| \le L_0\mbox{ for any } \boldsymbol{\beta}_i \in \partial f(\boldsymbol{x}) \mbox{ and } \boldsymbol{x} \in \Omega\bigg\}.
\end{equation}

In our ALCR, we minimize the sum of squared errors:
\begin{equation}\label{lse: augmented}
    \frac{1}{n} \sum_{i=1}^n (y_i - f(\boldsymbol{x}_i))^2
\end{equation}
over $f \in \mathcal{F}_A$. The infinite-dimensional problem (\ref{lse: augmented}) can be reduced to the following finite-dimensional convex program in the decision variables $f_1, \cdots, f_n \in \mathbb{R}$ and $\boldsymbol{\beta}_1, \cdots, \boldsymbol{\beta}_n \in \mathbb{R}^d$:
\begin{equation}\label{alcr}
\begin{aligned}
    \min \quad & \frac{1}{n}\sum_{i=1}^{n} (y_i - f_i)^2 &{}& \\
    \mbox{s.t.}\quad
    & f_j \geq f_i + \boldsymbol{\beta}_i^T (\boldsymbol{x}_j - \boldsymbol{x}_i) &\quad& i, j = 1, \cdots, n \\
    &\|\boldsymbol{\beta}_i-\boldsymbol{b}_0\| \leq L_0 &{}&  i = 1, \cdots, n.
\end{aligned}
\end{equation}
The solution $\hat{f}_{1, A}, \cdots, \hat{f}_{n, A}, \hat{\boldsymbol{\beta}}_{1, A}, \cdots, \hat{\boldsymbol{\beta}}_{n, A}$ to (\ref{alcr}) exists because (\ref{alcr}) is an optimization problem with continuous and coercive objective function over a nonempty convex subset (Proposition 7.3.1 and Theorem 7.3.7 in \citealp{kurdila2006convex}). Furthermore, the minimizing values $\hat{f}_{1, A}, \cdots, \hat{f}_{n, A}$ are unique because the objective function is strictly convex, but the $\hat{\boldsymbol{\beta}}_{i, A}$'s are not unique. We now define $\hat{f}_{n, A}: \Omega \rightarrow \mathbb{R}$ by
\begin{equation} 
\label{construction_of_ALCV_estimator}
    \hat{f}_{n, A}(\boldsymbol{x}) = \max_{1 \leq i \leq n}\{\hat{f}_{i, A} + \hat{\boldsymbol{\beta}}_{i, A}^T(\boldsymbol{x} - \boldsymbol{x}_i)\}
\end{equation}
for $\boldsymbol{x} \in \Omega$ and refer to $\hat{f}_{n, A}$ as the ALCR estimator.

The following theorem, Theorem \ref{theorem: alcr consistency}, establishes the strong uniform consistency of the ALCR estimator and its subgradients over the entire domain $\Omega$ as $n \rightarrow \infty$. The proof of Theorem \ref{theorem: alcr consistency} is provided in the supplementary material to this paper. 
\begin{theorem} \label{theorem: alcr consistency}
    (i) Assume \textbf{A1}, \textbf{A2}, \textbf{A3(i)}, \textbf{A4}, \textbf{A5}, and \textbf{A7}. Then, we have
\[\sup_{\boldsymbol{x} \in \Omega} |\hat{f}_{n, A} (\boldsymbol{x}) - f_0 (\boldsymbol{x})| \rightarrow 0\]
for all $n$ sufficiently large a.s.

    (ii) Assume \textbf{A1}, \textbf{A2}, \textbf{A3(i)}, \textbf{A4}--\textbf{A7}. Then, we have
\[\sup_{\boldsymbol{x} \in \Omega} \sup_{\boldsymbol{\beta} \in \partial \hat{f}_{n, A} (\boldsymbol{x})} \|\boldsymbol{\beta} - \nabla f_0 (\textbf{x})\| \rightarrow 0\]
for all $n$ sufficiently large a.s.
\end{theorem}

Theorem \ref{theorem: alcr consistency} ensures that the ALCR estimator and its subgradients converge uniformly on $\Omega$ to $f_0$ and $\nabla f_0$, respectively, with probability one as $n \rightarrow \infty$. This shows that the overfitting behavior is successfully eliminated in the ALCR estimator.

\begin{remark}
   It should be noted that the LCR estimator considered by \cite{mazumder2019a} is a special case of the ALCR estimator when $\boldsymbol{b}_0 = \boldsymbol{0}$. Thus, Theorem \ref{theorem: alcr consistency} establishes the strong uniform consistency of the LCR estimator and its subgradients over the entire domain $\Omega$.
\end{remark}

\begin{remark}\label{remark: beta0}
    The class $\mathcal{F}_A$ requires specifying $L_0$ and $\boldsymbol{b}_0$. In practice, $L_0$ can be determined by cross-validation. $\boldsymbol{b}_0$ can be obtained through data-driven methods. For example, one can use the slope of the linear regression estimator or some percentiles of the subgradients of the CR estimator. Alternatively, when a decision-maker has an estimate of the subgradient of $f_0$, one can use this estimate as $\boldsymbol{b}_0$. In the application context where one tries to estimate a cost function $f_0$, the subgradient of $f_0$ represents the marginal cost, and decision-makers can get an estimate of the marginal costs. In such a case, one can use this estimate as $\boldsymbol{b}_0$.
\end{remark}

\section{Weight-restricted convex regression}\label{sec: wrcr}
In this section, we propose the WRCR estimator by imposing explicit upper and lower bounds to the subgradients. This will help avoid very large values of the subgradients estimated from CR. 

Given $\boldsymbol{l}_0 \in \mathbb{R}^d$ and $\boldsymbol{u}_0 \in \mathbb{R}^d$, we consider the class of convex functions whose subgradients are bounded between $\boldsymbol{l}_0$ and $\boldsymbol{u}_0$:
\begin{equation}\label{class: weight}
    \mathcal{F}_B:= \bigg\{f: \Omega \rightarrow \mathbb{R} \:|\: f \mbox{ is convex and } \boldsymbol{l}_0 \leq \boldsymbol{\beta} \leq \boldsymbol{u}_0 \mbox{ for any }\boldsymbol{\beta} \in \partial f(\boldsymbol{x})  \mbox{ and }  \boldsymbol{x} \in   \Omega\bigg\}.
\end{equation}

In our proposed WRCR, we minimize the sum of squared errors:
\begin{equation}\label{lse: weight}
    \frac{1}{n} \sum_{i=1}^n (y_i - f(\boldsymbol{x}_i))^2
\end{equation}
over $f \in \mathcal{F}_B$.  The infinite-dimensional problem (\ref{lse: weight}) can be reduced to the following finite-dimensional convex program in the decision variables $f_1, \cdots, f_n \in \mathbb{R}$ and $\boldsymbol{\beta}_1, \cdots, \boldsymbol{\beta}_n \in \mathbb{R}^d$:
\begin{equation}\label{wrcr}
\begin{aligned}
    \min \quad & \frac{1}{n}\sum_{i=1}^{n} (y_i - f_i)^2 &{}& \\
    \mbox{s.t.}\quad
    & f_j \geq f_i + \boldsymbol{\beta}_i^T (\boldsymbol{x}_j - \boldsymbol{x}_i) &\quad& i, j = 1, \cdots, n \\
    &\boldsymbol{l}_0 \leq \boldsymbol{\beta}_i \leq \boldsymbol{u}_0 &{}&  i = 1, \cdots, n.
\end{aligned}
\end{equation}
The solution $\hat{f}_{1, B}, \cdots, \hat{f}_{n, B}, \hat{\boldsymbol{\beta}}_{1, B}, \cdots, \hat{\boldsymbol{\beta}}_{n, B}$ to (\ref{wrcr}) exists because (\ref{wrcr}) is an optimization problem with continuous and coercive objective function over a nonempty convex subset (Proposition 7.3.1 and Theorem 7.3.7 in \citealp{kurdila2006convex}). Furthermore, the minimizing values $\hat{f}_{1, B}, \cdots, \hat{f}_{n, B}$ are unique because the objective function is strictly convex, but the $\hat{\boldsymbol{\beta}}_{i, B}$'s are not unique. We now define $\hat{f}_{n, B}: \Omega \rightarrow \mathbb{R}$ by
\begin{equation} 
\label{construction_of_WRCV_estimator}
    \hat{f}_{n, B}(\boldsymbol{x}) = \max_{1 \leq i \leq n}\{\hat{f}_{i, B} + \hat{\boldsymbol{\beta}}_{i, B}^T(\boldsymbol{x} - \boldsymbol{x}_i)\}
\end{equation}
for $\boldsymbol{x} \in \Omega$ and refer to $\hat{f}_{n, B}$ as the WRCR estimator.

The following theorem, Theorem \ref{theorem: wrcr consistency}, establishes the strong uniform consistency of the WRCR estimator over the entire domain $\Omega$. The proof of Theorem \ref{theorem: wrcr consistency} is provided in the supplementary material to this paper.

\begin{theorem} \label{theorem: wrcr consistency}
(i) Assume \textbf{A1}, \textbf{A2}, \textbf{A3(i)}, \textbf{A4}, \textbf{A5}, and \textbf{A8}. Then, we have
$$\sup_{\boldsymbol{x} \in [0, 1]^d} |\hat{f}_{n, B} (\boldsymbol{x}) - f_0 (\boldsymbol{x})| \rightarrow 0$$
for all $n$ sufficiently large a.s.
    
(ii) Assume \textbf{A1}, \textbf{A2}, \textbf{A3(i)}, \textbf{A4}, \textbf{A5}, \textbf{A6}, and \textbf{A8}. Then, we have
\[\sup_{\boldsymbol{x} \in [0, 1]^d} \sup_{\boldsymbol{\beta} \in \partial \hat{f}_{n, B} (\boldsymbol{x})} \|\boldsymbol{\beta} - \nabla f_0 (\textbf{x})\| \rightarrow 0\]
for all $n$ sufficiently large a.s.
\end{theorem}

Theorem \ref{theorem: wrcr consistency} ensures that the WRCR estimator and its subgradients converge uniformly on $\Omega$ to $f_0$ and $\nabla f_0$, respectively, with probability one as $n \rightarrow \infty$. This shows that the overfitting behavior is successfully eliminated in the WRCR estimator.

\begin{remark}
\label{remark:wrcr and kuosmanen} It should be noted that the non-decreasing formulation proposed by \cite{kuosmanen2008representation} is a special case of the WRCR estimator when $\boldsymbol{l}_0 = \boldsymbol{0}$ and each component of $\boldsymbol{u}_0$ is $\infty$.
\end{remark}

\begin{remark}
\label{remark: dara-driven method for wrcr} 
The performance of the WRCR estimator relies on the proper choice of the lower bound $\boldsymbol{l}_0$ and upper bound $\boldsymbol{u}_0$. In the application context where the subgradient has a practical meaning, decision-makers can provide specific values of $\boldsymbol{l}_0$ and $\boldsymbol{u}_0$. For example, when $f_0$ represents a cost function, its subgradients represent the marginal costs, and decision-makers can have a sense of how high the marginal cost should be. Thus, they can provide specific values for $\boldsymbol{l}_0$ and $\boldsymbol{u}_0$. When such information is not available, one can rely on a data-driven method as follows. We first compute the subgradient $\hat{\boldsymbol{\beta}}_i = (\hat{\boldsymbol{\beta}}_i(1), \cdots, \hat{\boldsymbol{\beta}}_i(d))$ of the CR estimator for $i = 1, \cdots, n$ and let $\boldsymbol{l}_0(k)$ and $\boldsymbol{u}_0(k)$ be the $q$th and $(1-q)$th percentiles of $\{\hat{\boldsymbol{\beta}}_1(k), \cdots, \hat{\boldsymbol{\beta}}_n(k)\}$ for each $k \in \{1, \cdots, d\}$. We can compute $q$ using cross-validation, or decision-makers can provide the value of $q$ when they wish to exclude some unreliable results from the $\hat{\boldsymbol{\beta}}_i$'s. By letting $\boldsymbol{u}_0 = (\boldsymbol{u}_0(1), \cdots, \boldsymbol{u}_0(d))$ and $\boldsymbol{l}_0 = (\boldsymbol{l}_0 (1), \cdots, \boldsymbol{l}_0(d))$, we have data-driven estimates of the lower and upper bounds for WRCR.
\end{remark}

\section{Monte Carlo study}\label{sec: MC}

The goal of this section is to observe the numerical behavior of our proposed ALCR and WRCR estimators compared to some existing estimators. We start with Section \ref{subsec: illustration}, which illustrates the CR estimator's overfitting behavior numerically.

\subsection{Illustration of the overfitting behavior}
\label{subsec: illustration}

In this section, we observe the overfitting behavior of the CR estimator numerically. We consider $f_0:[0.2, 1.0] \rightarrow \mathbb{R}$ defined by $f_0(\boldsymbol{x}) = 1/\boldsymbol{x}$ for $\boldsymbol{x} \in [0.2, 1.0]$. The $\boldsymbol{x}_i$'s are evenly distributed, so we let $\boldsymbol{x}_i = 0.2 + 0.8i/n -0.8/(2n)$ for $i = 1, \cdots, n$. We generate the $y_i$'s from $y_i = f_0(\boldsymbol{x}_i) + \varepsilon_i$ for $i = 1, \cdots, n$, where the $\varepsilon_i$'s are i.i.d. normal random variables with mean 0 and variance 1. Once the $(\boldsymbol{x}_i, y_i)$'s are obtained, we compute the CR estimator $\hat{f}_n$ by solving (\ref{cr_new_formulation}) and (\ref{unique_beta_cr})  with Mosek.

To observe how the CR estimator behaves at the boundary of $\Omega = [0.2, 1.0]$, we compute  
\begin{equation}
    \label{max_abs_value}
\max_{1 \leq i \leq n} |\hat{f}_n(\boldsymbol{x}_i) - f_0(\boldsymbol{x}_i)|.
\end{equation}
We repeat this 100 times independently, generating 100 replications of (\ref{max_abs_value}), and use these 100 values to compute the 95\% confidence interval of $\mathbb{E}[\max_{1 \leq i \leq n} |\hat{f}_n(\boldsymbol{x}_i) - f_0(\boldsymbol{x}_i)|]$. Table \ref{tab: overfitting_cr_pcr} reports these 95\% confidence intervals for a wide range of $n$.

Another way to observe the overfitting behavior is by looking at the accuracy of the estimated subgradients. We thus compute 
\begin{equation}
    \label{max_abs_subgradient}
\max_{1 \leq i \leq n} |\mbox{subgrad }\hat{f}_n(\boldsymbol{x}_i) - \nabla f_0(\boldsymbol{x}_i)|,
\end{equation}
repeat this 100 times independently, generating 100 replications of (\ref{max_abs_subgradient}), and use these 100 values to compute the 95\% confidence interval of $\mathbb{E} [\max_{1 \leq i \leq n} |\mbox{subgrad }\hat{f}_n(\boldsymbol{x}_i) - \nabla f_0(\boldsymbol{x}_i)|]$. Table \ref{tab: overfitting_cr_pcr} reports these 95\% confidence intervals for a wide range of $n$.

The results in Table \ref{tab: overfitting_cr_pcr} show that the CR estimator produces inconsistent estimators of $f_0$ near the boundary of $\Omega$ as $n$ increases. It also generates subgradients whose magnitudes are increasing to infinity as $n \rightarrow \infty$. 

\begin{table}[ht]
\centering
\caption{95\% confidence intervals (CI's) of the maximum absolute error for the CR estimator and its subgradients.}
\begin{threeparttable}
    \begin{tabular}{c c c c }
    \hline
    & {CI's for $\mathbb{E} [\max_{1 \leq i \leq n} |\hat{f}_n(\boldsymbol{x}_i) - f_0 (\boldsymbol{x}_i)|]$} & {CI's for $\mathbb{E}[ \max_{1 \leq i \leq n} | \mbox{subgrad }\hat{f}_n(\boldsymbol{x}_i) - \nabla f_0 (\boldsymbol{x}_i)|]$}\\     
    \hline
    $n= 100$ & $0.85\pm0.10$ & $70.9\pm13.6$ \\
    $n= 200$ & $0.82\pm0.07$ & $115.2\pm19.0$ \\
    $n= 400$ & $0.94\pm0.06$ & $270.4\pm31.3$ \\
    % \midrule
    % \multicolumn{2}{c}{CI's for $\mathbb{E}[ \max_{1 \leq i \leq n} | \mbox{subgrad }\hat{f}_n(\boldsymbol{x}_i) - \nabla f_0 (\boldsymbol{x}_i)|]$} \\
    % \midrule
    %  $n= 100$ & $70.9\pm13.6$  \\
    %  $n= 200$ & $115.2\pm19.0$  \\
    %  $n= 400$ & $270.4\pm31.3$  \\
    \hline
    \end{tabular}
\end{threeparttable}
\label{tab: overfitting_cr_pcr}
\end{table}

\subsection{Comparisons of statistical performance}

This section is concerned with the numerical performance of our proposed ALCR and WRCR estimators compared with three existing estimators, CR, PCR, and LCR estimators. The CR estimator is defined in (\ref{construction_of_CV_estimator}). The LCR estimator is a special case of the ALCR estimator (\ref{construction_of_ALCV_estimator}) when $\boldsymbol{b}_0 = \boldsymbol{0} \in \mathbb{R}^d$. 

\subsubsection{Setup}\label{subsec: setup}

Consider the following two test functions:
\begin{itemize}
    \item[] Type A: \(f_0(\boldsymbol{x}) = 0.1x(1)+0.1x(2)+0.1x(3)+0.3(x(1)x(2)x(3))^{1/3}\)
    \item[] Type B: \(f_0(\boldsymbol{x}) = \prod_{p=1}^{d} x(p)^{\frac{0.8}{d}}\),
\end{itemize}
where $y_i = f_0(\boldsymbol{x}_i)+\varepsilon_i$ for $i=1,\cdots,n$. We generate the $\boldsymbol{x}_i$'s, independently of one another, from the uniform distribution over $[1, 10]^d$ and draw the $\varepsilon_i$'s from the normal distribution with mean 0 and variance $\sigma^2$, where $\sigma^2$ is determined by the signal-to-noise ratio (SNR), $\text{SNR}= \text{Var}[f_0 (\boldsymbol{x}_1)]/\sigma^2$. 

In all simulations, we generate $2n$ independent observations and equally split them into the training and validation sets.  We first use the training set to compute the CR estimator and its subgradients $\hat{\boldsymbol{\beta}}_i$'s by solving (\ref{cr_new_formulation}). Next, we use the validation set to compute $\boldsymbol{b}_0$ and all the tuning parameters. We find $\boldsymbol{b}_0$ by computing the slope of the linear regression estimator with the validation set. The tuning parameters we need to compute from the validation set are $L$ for the LCR estimator, $\lambda_n$ for the PCR estimator, $L_0$ for the ALCR estimator, and $q$ (defined in Remark \ref{remark: dara-driven method for wrcr}) for the WRCR estimator. We use the 5-fold cross-validation method to compute these tuning parameters. For example, to find $\lambda_n$ for the PCR estimator using the 5-fold cross-validation method, we split the validation set into five equally sized sets, say $S_1, S_2, \cdots, S_5$. For each $\lambda \in \mathbb{R}$, we define the following cross-validation function:
\[\mbox{CV}(\lambda) \triangleq \frac{1}{5} \sum_{k = 1}^5 \sum_{(\boldsymbol{x}_j, y_j) \in S_k} (y_j - f_{\lambda, n}^{[k]}(\boldsymbol{x}_j))^2,\]
where $f_{\lambda, n}^{[k]}$ is the PCR estimator in (\ref{pcr}) and (\ref{construction_of_PCV_estimator}) computed with $\lambda_n = \lambda$ and the $(\boldsymbol{x}_i, y_i)$'s replaced by $\{(\boldsymbol{x}_j, y_j) \;| \; (\boldsymbol{x}_j, y_j) \mbox{ is in the validation set, but }  (\boldsymbol{x}_j, y_j) \notin S_k\}$. We then evalulate $\mbox{CV}(\lambda)$ for each value $\lambda$ in the set of candidate values, say $\{\lambda_1, \cdots, \lambda_m\}$, and select the value with the least value of $\mbox{CV}(\lambda)$. We find $L$, $L_0$, and $q$ using the 5-fold cross-validation method in a similar fashion. For the set of candidate values, we use  $\{0.01, 0.02, \cdots, 0.49\}$ to search for $q$. To search for $L_0$, we use 50 equally spaced values from $[0,\overline{L}]$, where $\overline{L} = \max_{1 \leq i \leq n}\|\hat{\boldsymbol{\beta}}_i - \boldsymbol{b}_0\|$. For $L$, we use 50 equally spaced values from $[0.1, 5.0]$.

In the following experiments, we use the standard solver Mosek (9.2.44) within the Julia/JuMP package to compute optimization problems. All computations are performed on a computing cluster with Xeon @2.8 GHz processors, 10 CPUs, and 8 GB RAM per CPU.

\subsubsection{Statistical performance}

We first consider the Type A function with $n=\{50, 100, 200\}$ and $\text{SNR}=3$ to compare our proposed estimators to the CR, PCR, and LCR estimators.

As described in Section \ref{subsec: setup}, we generate the training and validation sets and use them to compute all the tuning parameters. Next, we generate 1000 independent replications of $(\boldsymbol{x}_1, y_1)$, say $((\boldsymbol{x}_i, y_i): 1 \leq i \leq 1000)$ and use the data along with the tuning parameters to compute prediction errors of the LCR, PCR, ALCR, and WRCR estimators. 

To measure the accuracy of the ALCR estimator $\hat{f}_{n, A}$, we compute the empirical mean square error (MSE) value as follows:
\begin{equation}
\label{MSE}
\frac{1}{1000}\sum_{i = 1}^{1000}(\hat{f}_{n, A}(\boldsymbol{x}_i) - f_0(\boldsymbol{x}_i))^2.
\end{equation}
We repeat this 50 times independently, generating 50 replications of (\ref{MSE}). Table \ref{tab: beta} reports the 95\% confidence intervals of the expected MSE that are computed using these 50 values for various values of $n$. We repeat this procedure for each of the PCR, LCR, and WRCR estimators and report the confidence intervals of their expected MSEs in Table \ref{tab: beta}.

\begin{table}[ht]
\centering
\caption{95\% confidence interval of the expected MSE using Type A function.}
\begin{threeparttable}
    \begin{tabular}{lc c c c c}
    \hline
     &CR &PCR &LCR &ALCR &WRCR\\
    \hline
$n = 50$ & $1.231\pm0.889$ & $0.085\pm 0.014$ & $0.073\pm0.008$ & $0.036\pm0.006$ & $0.045\pm0.007$\\
$n = 100$ &  $1.019\pm1.140$ & $0.046\pm0.005$ & $0.033\pm0.004$ & $0.019\pm0.002$ & $0.022\pm0.002$\\
$n = 200$  & $1.179\pm1.168$ & $0.026\pm 0.002$ & $0.019\pm0.002$ & $0.013\pm0.001$ & $0.015\pm0.001$\\
    \hline
    \end{tabular}
\end{threeparttable}
\label{tab: beta}
\end{table}

Although all of the PCR, LCR, ALCR, and WRCR estimators seem to benefit from the additional restrictions on subgradients, our ALCR and WRCR estimators yield more significant improvements in the predictive power over the PCR and LCR estimators. 

\begin{figure}[ht]
    \centering
    \includegraphics[width=0.8\textwidth]{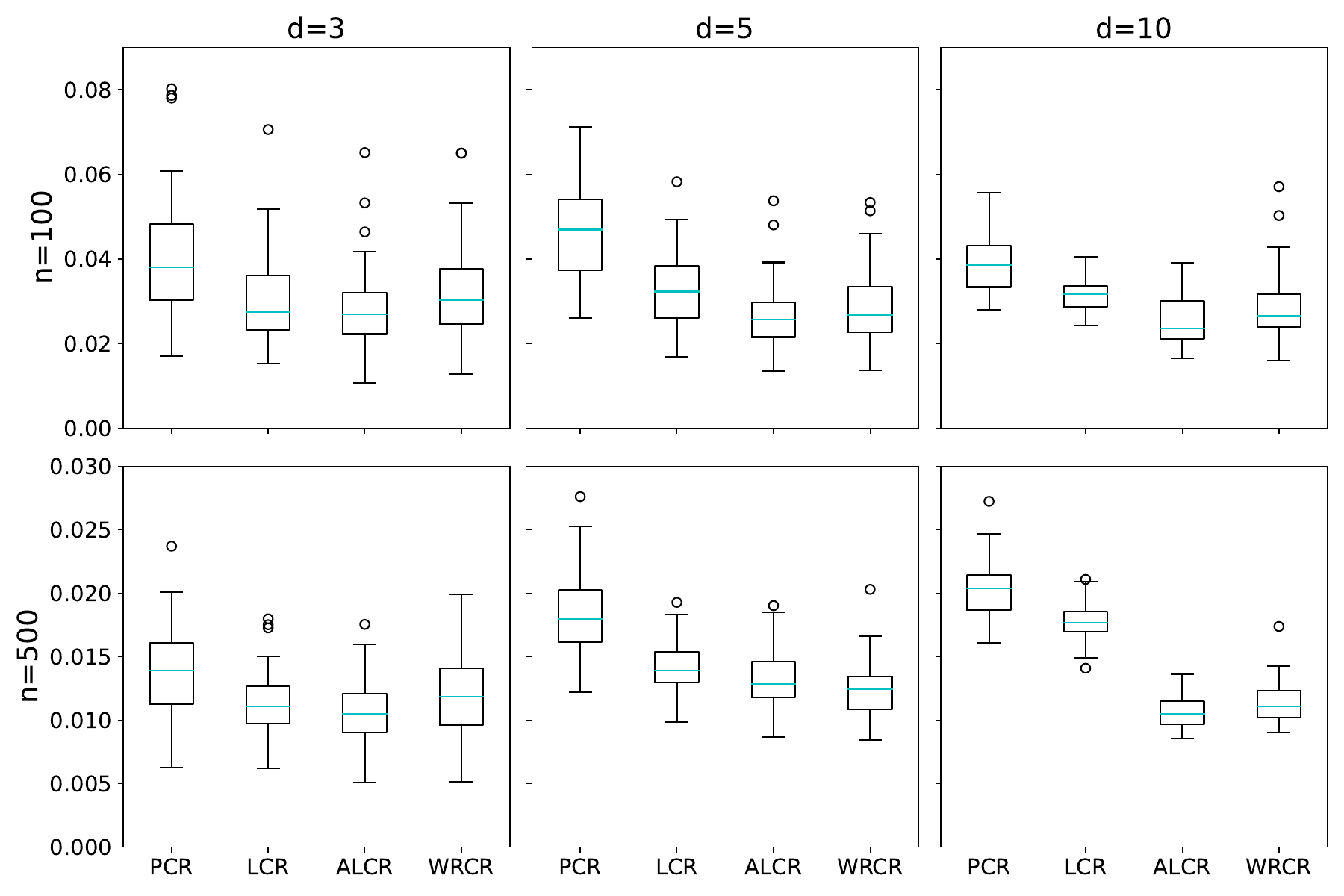}
    \caption{Each box plot is created with 50 MSE values for each of the PCR, LCR, ALCR, and WRCR estimators for difference values of $n$ and $d$. We use the Type B function.}
    \label{fig: mse_prod}
\end{figure}

We next repeat this procedure with the Type B function, generating 50 replications of the MSE value for each of the LCR, PCR, ALCR, and WRCR estimators, and create box plots each with 50 MSE values. In Figure \ref{fig: mse_prod}, each box plot shows the median (in the middle of each box), the first quantile (lower end of each box), the third quartile (higher end of each box), and outliers as small circles. Figure \ref{fig: mse_prod} shows the box plots for a wide range of $d$ when $n = 100$ and $n = 500$. Our proposed estimators, the ALCR and WRCR estimators, perform better than other estimators in most cases, especially when the dimension is higher, e.g., $d = 10$.

\begin{figure}[ht]
    \centering
    \includegraphics[width=0.9\textwidth]{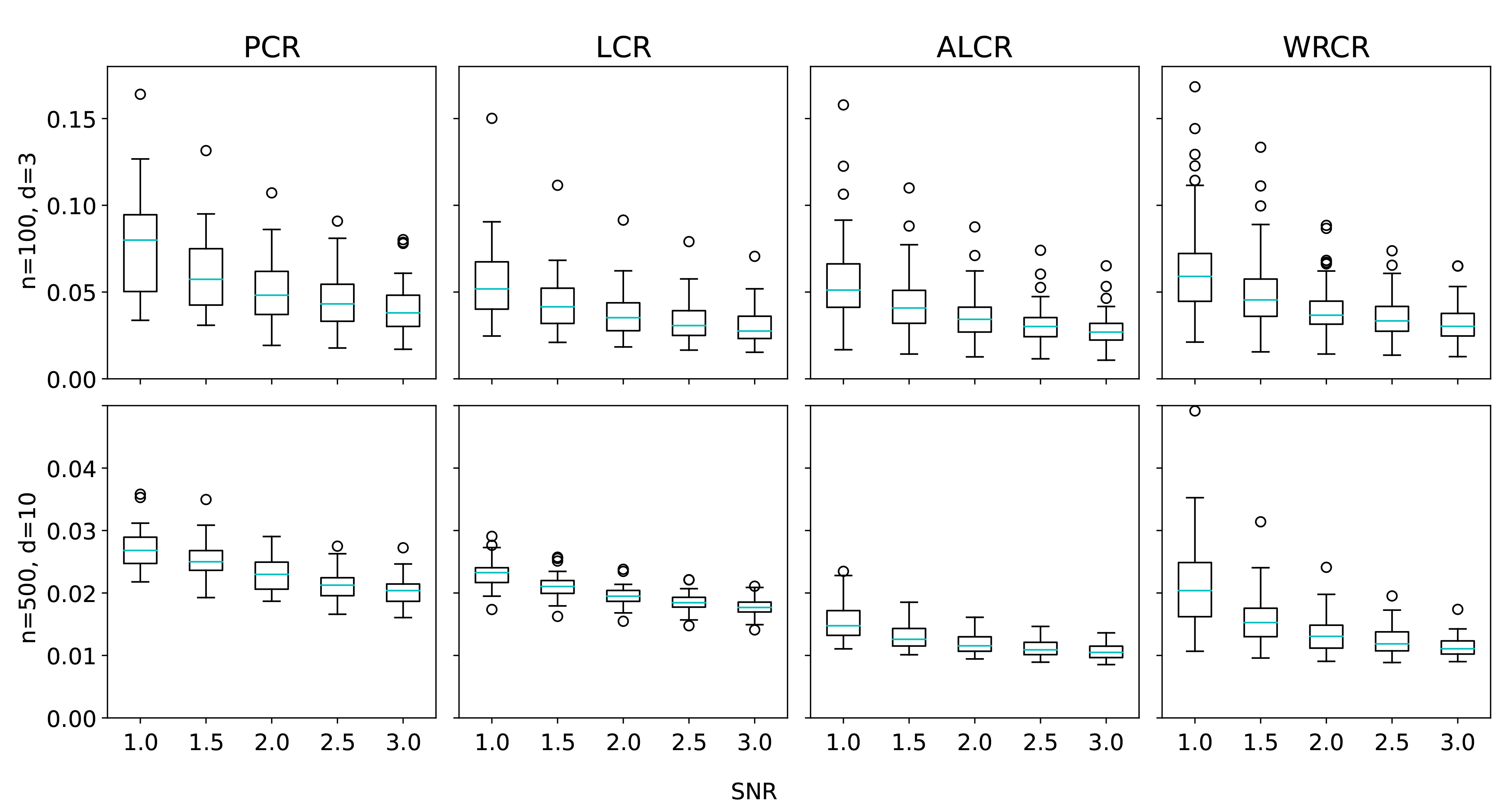}
    \caption{Each box plot is created with 50 MSE values for each of the PCR, LCR, ALCR, and WRCR estimators for different values of SNR. We use the Type B function.}
    \label{fig: mse_snr}
\end{figure}

We next use the Type B function to explore how the noise level affects the predictive performance of the proposed estimators. We repeat the previous experiment with the Type B function. Figure \ref{fig: mse_snr} reports the box plots each with 50 MSE values for a wide range of SNR$\in\{1.0, 1.5, 2.0, 2.5, 3.0\}$ when $n = 100, d = 3$ and when $n=500, d = 10$.

LCR performs similarly to ALCR when the sample size is small ($n=100, d=3$). However, the new method shows better predictive power in large sample settings ($n=500, d=10$). In these settings, WRCR also performs reasonably well compared to other benchmarks but does not show such an advantage at high noise levels. The supplementary material to this paper provides additional numerical results with different underlying functions $f_0$. 

\section{Regulation on electricity distribution firms in Finland}\label{sec: application}

In Finland, electricity distribution firms typically enjoy a natural local monopoly due to prohibitively expensive construction fees. This forces governments to establish regulatory agencies to monitor the electricity distribution firms, reduce their local monopoly power, and provide incentives to those adopting the best technology. The EV is one of the pioneers in practically implementing energy regulation programs, which involve cost-efficiency analysis for distribution firms. The traditional frontier estimation techniques, such as data envelopment analysis and stochastic frontier analysis, were applied by EV from 2005 to 2011, and a more reliable CR model was adopted from 2012 onward \citep{kuosmanen2012stochastic}. Recently, EV considered developing new techniques to reduce the overfitting behavior in CR and increase the accuracy of its incentive regulatory model for Finnish electricity distribution firms \citep{kuosmanen2022}. In our analysis, we examine the performance of our proposed estimators on the cost frontier estimation problem for Finnish electricity distribution firms.

\subsection{Frontier estimation model}
\label{subsec: app beta}

We consider a general cost frontier model in production economics \citep{kuosmanen2020conditional}:
\begin{equation}
\label{cost function}
    y_i = C(\boldsymbol{x}_i) + \varepsilon_i
\end{equation}
for $i = 1, \cdots, n$, where $C$ is a nondecreasing and convex cost function, and the $\varepsilon_i$'s are the error terms representing random effects. $\boldsymbol{x}_i$ is the vector of performance variables for firm $i$, and $y_i$ is the variable cost of firm $i$. The variables are specified as follows:
\begin{eqnarray*}
y & = &\mbox{Variable cost representing the controllable operational expenditure (KOPEX, \euro{})}\\
x(1) & = & \mbox{Energy supply (GWh, weighted by voltage)}\\
x(2) & = & \mbox{Length of the network (km)}\\
x(3) & = & \mbox{Number of user points}\\
x(4) & = &  \mbox{Fixed cost representing the capital stock (regulatory asset value, NKA, \euro{})}
\end{eqnarray*}
We collect data from the EV over three regulation periods with 86 Finnish electricity distribution firms, covering the years from 2008 to 2020. The total number of $(\boldsymbol{x}_i, y_i)$ pairs in our data set is $1042$, with some companies going bankrupt after 2010. The original dataset is applied to build the incentive regulation model for Finnish electricity distribution firms \citep{kuosmanen2022}, and its earlier version has been widely used; see \cite{kuosmanen2012stochastic}, \cite{kuosmanen2013what}, and \cite{kuosmanen2020conditional} for details.

To estimate the cost function $C$ in \eqref{cost function}, we consider the following variant of CR that incorporates the monotonicity of $C$: 
\begin{equation}\label{cnls} 
\begin{aligned}
    \min_{\boldsymbol{\beta}, \alpha, \varepsilon} \quad & \frac{1}{n}\sum_{i=1}^{n} \varepsilon_i^2 &{\quad}& \\
    \mbox{\textit{s.t.}}\quad
    & y_i = \alpha_i + \boldsymbol{\beta}_i^T \boldsymbol{x}_i + \varepsilon_i  &{}&  i = 1, \cdots, n\\
    &\alpha_i + \boldsymbol{\beta}_i^T \boldsymbol{x}_i  \geq \alpha_j + \boldsymbol{\beta}_j^T \boldsymbol{x}_i  &{}&   i, j = 1, \cdots, n  \\
    &\boldsymbol{\beta}_i\geq \boldsymbol{0} &{}&   i = 1, \cdots, n.  
\end{aligned}
\end{equation}
The last constraint in \eqref{cnls}  guarantees that the estimated cost function is non-decreasing. The monotonicity assumption on production/cost function has been identified in various areas such as operations research and economics; see, e.g.,  \cite{kuosmanen2010data} and \cite{kuosmanen2012stochastic2}.  However, the additional monotonicity constraint in  \eqref{cnls} does not eliminate the overfitting behavior. 

Using our data set $((\boldsymbol{x}_i, y_i): 1 \leq i \leq 1042)$, we compute the CR estimator and its subgradients $\hat{\boldsymbol{\beta}}_i = (\hat{\beta}_i(1), \hat{\beta}_i(2), \hat{\beta}_i(3), \hat{\beta}_i(4))$ at $\boldsymbol{x}_i$. Table \ref{tab: beta-app} reports some summary statistics, such as the mean, standard deviation, minimum, 10th percentile, median, 90th percentile, and maximum, for $(\hat{\beta}_i(k): 1 \leq i \leq 1042)$ for each  $k \in \{1, 2, 3, 4\}$.

\begin{table}[ht]
\centering
\caption{Summary of the subgradients of the CR estimator with the Finnish electricity distribution firm data.}
\begin{threeparttable}
    \begin{tabular}{l r r r r}%p{2.3cm}m{3.5cm}m{3cm}m{3.2cm}r{1.6cm}}
    \hline
         & $\hat{\beta}_i(1)$ & $\hat{\beta}_i(2)$ & $\hat{\beta}_i(3)$ & $\hat{\beta}_i(4)$  \\
                  & (\euro{} cents/kWh) &  (\euro{} /km) &  (\euro{} /customer) &  \\
    \hline
    Mean & 2.7 & 0.6 & 66.8 & 27.1 \\
    Standard deviation  & 4.6 & 1.7 & 627.2 & 107.6 \\
    Minimum & 0.0 & 0.0 & 0.0 & 0.0 \\
    10th percentile   & 0.0 & 0.0 & 0.1 & 0.1 \\
    Median & 1.8 & 0.5 & 34.3 & 9.1\\
    90th percentile   & 6.4 & 0.8 & 75.5 & 41.4 \\
    Maximum & 69.4 & 39.5 & 19887.3 & 2385.4 \\
    \hline
    \end{tabular}
\end{threeparttable}
\label{tab: beta-app}
\end{table}

For each variable, the range between the minimum and maximum estimated subgradients is quite large. Those large subgradients usually occur near the boundary of the domain and, hence, lead to the overfitting behavior in the CR estimator. The overfitting problem may yield invalid results in frontier estimation and produce misleading benchmarks for the acceptable level of cost in future regulation periods. 

The 90th percentiles indicate that the estimated subgradients within the 90th percentile should not be very large. We notice that the changes in the estimated subgradients are much smaller within the range between the 10th and 90th percentiles compared to the changes in the range between the minimum and the maximum. This difference is caused by a few big firms with extreme data values that lead to overfitting behavior in the estimation of the cost function using CR. 

The descriptive statistics in Table \ref{tab: beta-app} provide some information that we can use in our proposed estimators. For example, the median values in Table \ref{tab: beta-app} are natural candidates for $\boldsymbol{b}_0$ in the ALCR estimator.

\subsection{Evaluating the predictive performance}

For the next regulation period between 2024 and 2031, improving the predictive accuracy of the cost frontier model is one of the main concerns of EV regulators. In this section, we compare the two proposed estimators with several key benchmark methods that have been used on the Finnish electricity distribution firm data. For this purpose, we need to slightly modify the CR, PCR, ARCR, and WRCR estimators. The CR estimator \eqref{cnls} is a variant of the CR estimator where the monotonicity of the underlying function is incorporated as $\boldsymbol{\beta}_i \geq 0$ for $i = 1, \cdots, n$ in the constraint. We also add these conditions to the constraints of PCR's, LCR's, ARCR's, and WRCR's formulations. 

In all computations, we use the observations in the first 9 years $((\boldsymbol{a}_i, b_i): i = 1, \cdots, 734)$  as the training data set and the observations in the last 4 years $((\boldsymbol{c}_i, d_i): i = 1, \cdots, 308)$ as the test data set. We set $\boldsymbol{b}_0 = (1.8, 0.5, 34.3, 9.1)$ as the median values reported in Table \ref{tab: beta-app}. We use the training data set to compute all other tuning parameters, $\lambda$ in PCR, $L$ in LCR, $L_0$ in ARCR, and $q$ (defined in Remark \ref{remark: dara-driven method for wrcr}) in WRCR. We use the 5-fold cross-validation method with the following set of candidate values: 50 equally spaced values from $[1, 500]$ for each of $\lambda_n$, $L$ and $L_0$, and $\{0.01,0.02, \cdots, 0.49\}$ for $q$. Next, we use the test data set to measure the accuracy of each estimator. For example, to measure the accuracy of the CR estimator, we define in-sample root mean square error (in-sample RMSE) as follows:
\[\left\{\frac{1}{734}\sum_{i = 1}^{734} (b_i - \mbox{ CR at }\boldsymbol{a}_i)^2\right\}^{1/2}\]
and out-of-sample root mean square error (out-of-sample RMSE) as follows:
\[\left\{\frac{1}{308}\sum_{i = 1}^{308} (d_i - \mbox{CR at }\boldsymbol{c}_i)^2\right\}^{1/2}.\]

Table \ref{tab: rmse} reports the in-sample and out-of-sample RMSEs for various estimators. 
\begin{table}[ht]
\centering
\caption{In-sample and out-of-sample RMSEs.}
\begin{threeparttable}
    \begin{tabular}{l r r r r r}
    \hline
    &CR &PCR & LCR &ALCR &WRCR\\
    \hline
    In-sample RMSE     &1133.1 & 1384.4 & 1720.7& 1619.1 & 1626.0\\
    Out-of-sample RMSE &111255.4 & 1775.1 & 1748.4& 1716.0 & 1644.9\\ 
    \hline
    \end{tabular}
\end{threeparttable}
\label{tab: rmse}
\end{table}

The CR estimator results in a severe overfitting problem, where the in-sample RMSE is very small while the out-of-sample RMSE appears to be very large. The PCR, LCR, and our proposed estimators (ALCR and WRCR) yield results that are significantly better than those of  CR, demonstrating that an additional bound on the subgradients helps improve the predictive performance of CR. We observe that the proposed estimators work best in terms of the prediction accuracy. The WRCR estimator achieves the smallest out-of-sample RMSEs, indicating that it outperforms other CR estimators in terms of predictive power.

From the regulator's perspective, it can be hard to interpret the tuning parameter $\lambda_n$ in PCR and $L$ in LCR. This makes the model hard to be trusted by practitioners. By contrast, the tuning parameter $L_0$ in ALCR can be interpreted as the distance between the estimated marginal costs and the reference vector $\boldsymbol{b}_0$ (the median values of the subgradients of the CR estimator). The $q$th and $(1-q)$th percentiles used in the WRCR can be seen as the proportion of extreme values obtained from CR that the regulators would like to exclude from the subsequent estimation.

\section{Conclusions}\label{sec: conclusion}
In this paper,  we provide theoretical evidence of the overfitting behavior of the CR estimator near the boundary of the domain. To eliminate the overfitting behavior, we consider placing bounds on the subgradients of the fitted convex function. In particular, we propose two practical ways to place the bounds, leading our proposed estimators, the ALCR and WRCR estimators. Monte Carlo studies have shown the superior performance of our proposed estimators.

This work was initially motivated by the observed overfitting problem in the regulatory models previously used by the Finnish EV and the need to improve the prediction accuracy of their models in the 5th and 6th regulation periods from 2024 to 2031. The two newly proposed estimators allow EV to have a model that does not show the overfitting behavior and that outperforms the existing CR estimators.

There are several open research avenues. In this paper, we conducted numerical experiments in the cases where $d\ll n$. Understanding and investigating the behavior of the CR estimator when the number of variables is large could be a direction of interest. Furthermore, the predictive performance of our approaches relies on the choice of tuning parameters. It would be interesting to develop a tuning-free CR model that does not show overfitting behavior.

%%%%%%%%%%%%%%%%%
\section*{Acknowledgements}

The authors acknowledge the computational resources provided by the Aalto Science-IT project. Zhiqiang Liao gratefully acknowledges financial support from the HSE Support Foundation [grant no. 18-3419] and the Jenny and Antti Wihuri Foundation [grant no. 00220201]. Sheng Dai gratefully acknowledges financial support from the OP Group Research Foundation [grant no.~20230008].

\bibliographystyle{agsm.bst}

\bibliography{References}

@techreport{kuosmanen2022,
	title        = {{Kohtuullinen muuttuva kustannus sähkön jakeluverkkoyhtiöiden valvontamallissa: Ehdotus tehostamiskannustimen kehittämiseksi 6. ja 7. valvontajaksoilla vuosina 2024--2031 (in Finnish)}},
	author       = {Kuosmanen, Timo and Kuosmanen, Natalia and Dai, Sheng},
	year         = 2022,
	note         = {Available from: \url{energiavirasto.fi}, accessed 03.18.2023}
}

@book{kurdila2006convex,
	title        = {Convex functional analysis},
	author       = {Kurdila, Andrew J and Zabarankin, Michael},
	year         = 2006,
	publisher    = {Springer Science \& Business Media},
	address      = {Switzerland}
}

@inproceedings{Groeneboom1996,
	title        = {Inverse problems in statistics},
	author       = {P. Groeneboom},
	year         = 1996,
	booktitle    = {Proceedings of the St. Flour Summer School in Probability. Lecture Notes in Math.},
	publisher    = {Springer},
	address      = {Berlin},
	series       = 1648,
	pages        = {67--164}
}

@article{Bronshtein1976,
	title        = {$\epsilon$-entropy of convex sets and functions},
	author       = {E. M. Bronshtein},
	year         = 1976,
	journal      = {Siberian Math. J.},
	volume       = 17,
	pages        = {393--398}
}

@article{LuoLim2016,
	title        = {On consistency of least absolute deviations estimators of convex functions},
	author       = {Y. Luo and E. Lim},
	year         = 2016,
	journal      = {International Journal of Statistics and Probability},
	volume       = 5,
	number       = 2,
	pages        = {1--18}
}

@book{Rockafella1997,
	title        = {Convex Analysis},
	author       = {R. R. Rockafellar},
	year         = 1997,
	publisher    = {Princeton University Press}
}

@article{lever2016points,
	title        = {Points of significance: model selection and overfitting},
	author       = {Lever, Jake and Krzywinski, Martin and Altman, Naomi},
	year         = 2016,
	journal      = {Nature Methods},
	publisher    = {Nature Publishing Group},
	volume       = 13,
	pages        = {703--705}
}

@article{podinovski2016optimal,
	title        = {Optimal weights in DEA models with weight restrictions},
	author       = {Podinovski, Victor V},
	year         = 2016,
	journal      = {European Journal of Operational Research},
	publisher    = {Elsevier},
	volume       = 254,
	pages        = {916--924}
}

@inproceedings{blanchet2019multivariate,
	title        = {Multivariate distributionally robust convex regression under absolute error loss},
	author       = {Blanchet, Jose and Glynn, Peter W and Yan, Jun and Zhou, Zhengqing},
	year         = 2019,
	booktitle    = {Advances in Neural Information Processing Systems},
	volume       = 32
}

@article{lim2021consistency,
	title        = {Consistency of penalized convex regression},
	author       = {Lim, Eunji},
	year         = 2021,
	journal      = {International Journal of Statistics and Probability},
	publisher    = {Canadian Center of Science and Education},
	volume       = 10,
	pages        = {1--69}
}

@article{kuosmanen2020conditional,
	title        = {Conditional yardstick competition in energy regulation},
	author       = {Kuosmanen, Timo and Johnson, Andrew L},
	year         = 2020,
	journal      = {The Energy Journal},
	publisher    = {International Association for Energy Economics},
	volume       = 41,
	pages        = {67--92}
}

@article{liao2024convex,
	title        = {Convex support vector regression},
	author       = {Liao, Zhiqiang and Dai, Sheng and Kuosmanen, Timo},
	year         = 2024,
	journal      = {European Journal of Operational Research},
	publisher    = {Elsevier},
	volume       = 313,
	pages        = {858--870}
}

@article{ghosal2017univariate,
	title        = {On univariate convex regression},
	author       = {Ghosal, Promit and Sen, Bodhisattva},
	year         = 2017,
	journal      = {Sankhya A},
	publisher    = {Springer},
	volume       = 79,
	pages        = {215--253}
}

@article{afriat1972efficiency,
	title        = {Efficiency estimation of production functions},
	author       = {Afriat, S N},
	year         = 1972,
	journal      = {International Economic Review},
	volume       = 13,
	pages        = {568--598}
}

@inproceedings{balazs2015near,
	title        = {Near-optimal max-affine estimators for convex regression},
	author       = {Bal{\'{a}}zs, G{\'{a}}bor and Gy{\"{o}}rgy, Andr{\'{a}}s and Szepesv{\'{a}}ri, Csaba},
	year         = 2015,
	booktitle    = {18th Artificial Intelligence and Statistics},
	publisher    = {PMLR},
	pages        = {56--64}
}

@article{banker1993maximum,
	title        = {Maximum likelihood, consistency and data envelopment analysis. A statistical foundation},
	author       = {Banker, Rajiv D.},
	year         = 1993,
	journal      = {Management Science},
	volume       = 39,
	pages        = {1265--1273}
}

@article{bertsimas2021sparse,
	title        = {Sparse convex regression},
	author       = {Bertsimas, Dimitris and Mundru, Nishanth},
	year         = 2021,
	journal      = {INFORMS Journal on Computing},
	volume       = 33,
	pages        = {262--279}
}

@article{hannah2013multivariate,
	title        = {Multivariate convex regression with adaptive partitioning},
	author       = {Hannah, Lauren A and Dunson, David B},
	year         = 2013,
	journal      = {Journal of Machine Learning Research},
	volume       = 14,
	pages        = {3153--3188}
}

@article{hildreth1954point,
	title        = {Point estimates of ordinates of concave functions},
	author       = {Hildreth, Clifford},
	year         = 1954,
	journal      = {Journal of the American Statistical Association},
	volume       = 49,
	pages        = {598--619}
}

@article{kuosmanen2008representation,
	title        = {Representation theorem for convex nonparametric least squares},
	author       = {Kuosmanen, Timo},
	year         = 2008,
	journal      = {Econometrics Journal},
	volume       = 11,
	pages        = {308--325}
}

@article{kuosmanen2010data,
	title        = {Data envelopment analysis as nonparametric least-squares regression},
	author       = {Kuosmanen, Timo and Johnson, Andrew L.},
	year         = 2010,
	journal      = {Operations Research},
	volume       = 58,
	pages        = {149--160}
}

@article{kuosmanen2012stochastic,
	title        = {Stochastic semi-nonparametric frontier estimation of electricity distribution networks: Application of the StoNED method in the Finnish regulatory model},
	author       = {Kuosmanen, Timo},
	year         = 2012,
	journal      = {Energy Economics},
	volume       = 34,
	pages        = {2189--2199}
}

@article{kuosmanen2012stochastic2,
	title        = {Stochastic non-smooth envelopment of data: Semi-parametric frontier estimation subject to shape constraints},
	author       = {Kuosmanen, Timo and Kortelainen, Mika},
	year         = 2012,
	journal      = {Journal of Productivity Analysis},
	volume       = 38,
	pages        = {11--28}
}

@article{kuosmanen2013what,
	title        = {What is the best practice for benchmark regulation of electricity distribution? Comparison of DEA, SFA and StoNED methods},
	author       = {Kuosmanen, Timo and Saastamoinen, Antti and Sipil{\"{a}}inen, Timo},
	year         = 2013,
	journal      = {Energy Policy},
	volume       = 61,
	pages        = {740--750}
}

@article{kuosmanen2021shadow,
	title        = {Shadow prices and marginal abatement costs: Convex quantile regression approach},
	author       = {Kuosmanen, Timo and Zhou, Xun},
	year         = 2021,
	journal      = {European Journal of Operational Research},
	volume       = 289,
	pages        = {666--675}
}

@article{lee2013a,
	title        = {A more efficient algorithm for convex nonparametric least squares},
	author       = {Lee, Chia Yen and Johnson, Andrew L. and Moreno-Centeno, Erick and Kuosmanen, Timo},
	year         = 2013,
	journal      = {European Journal of Operational Research},
	volume       = 227,
	pages        = {391--400}
}

@article{lim2012consistency,
	title        = {Consistency of multidimensional convex regression},
	author       = {Lim, Eunji and Glynn, Peter W.},
	year         = 2012,
	journal      = {Operations Research},
	volume       = 60,
	pages        = {196--208}
}

@article{mazumder2019a,
	title        = {A computational framework for multivariate convex regression and its variants},
	author       = {Mazumder, Rahul and Choudhury, Arkopal and Iyengar, Garud and Sen, Bodhisattva},
	year         = 2019,
	journal      = {Journal of the American Statistical Association},
	volume       = 114,
	pages        = {318--331}
}

@article{seijo2011nonparametric,
	title        = {Nonparametric least squares estimation of a multivariate convex regression function},
	author       = {Seijo, Emilio and Sen, Bodhisattva},
	year         = 2011,
	journal      = {The Annals of Statistics},
	volume       = 39,
	pages        = {1633--1657}
}

@article{yagi2020shape,
	title        = {{Shape-constrained kernel-weighted least squares: Estimating production functions for Chilean manufacturing industries}},
	author       = {Yagi, Daisuke and Chen, Yining and Johnson, Andrew L. and Kuosmanen, Timo},
	year         = 2020,
	journal      = {Journal of Business \& Economic Statistics},
	volume       = 38,
	pages        = {43--54}
}

%-----------------%
% 
%-----------------%

\clearpage
\newpage
\baselineskip 20pt
\section*{Appendix}\label{sec:app}
\captionsetup[figure]{labelfont={bf},labelformat={default},labelsep=period,name={Fig.}}

%-----------------%
% 
%-----------------%

\renewcommand{\thesubsection}{\Alph{subsection}}
\renewcommand{\thetable}{B\arabic{table}}
\setcounter{table}{0}
\renewcommand{\theequation} {A.\arabic{equation}}
\setcounter{equation}{0}

\subsection{Proofs}

To prove Theorem \ref{theorem: inconsistency}, we need Lemma \ref{lemma:01} and Lemma \ref{lemma:02}, which are proved in Sections A.1 and A.2.
\begin{lemma}
\label{lemma:01}
Consider the problem of minimizing
\begin{equation}
\label{eqn:01} \sum_{i = 1}^n (y_i + \alpha_0 + \boldsymbol{\beta}_0^T\boldsymbol{x}_i - g(\boldsymbol{x}_i))^2
\end{equation}
over $g \in \mathcal{F}$, for given $\alpha_0 \in \mathbb{R}$ and $\boldsymbol{\beta}_0 \in \mathbb{R}^d$. Then, $\hat{g}_n:\Omega \rightarrow \mathbb{R}$ defined by $\hat{g}_n (\boldsymbol{x}) = \hat{f}_n(\boldsymbol{x}) + \alpha_0 + \boldsymbol{\beta}_0^T \boldsymbol{x}$ for $\boldsymbol{x} \in \Omega$ is a solution to (\ref{eqn:01}).
\end{lemma}

\begin{lemma}
\label{lemma:02}
 Assume \textbf{A1}--\textbf{A6}. Suppose $\nabla f_0 (\boldsymbol{0}) < \boldsymbol{0}$. Then, there exists $\epsilon_0 > 0$ such that for any $\epsilon < \epsilon_0$,
\[\liminf_{n \rightarrow \infty} \mathbb{P} (\hat{f}_n(\boldsymbol{0}) > f_0(\boldsymbol{0}) + \epsilon) > 0.\]
\end{lemma}

\subsubsection{Proof of Lemma \ref{lemma:01}}
\begin{proof}
Obviously, $\hat{g}_n$ belongs to $\mathcal{F}$. Now, we will show $\hat{g}_n$ is a minimizer of (\ref{eqn:01}). For any $g \in \mathcal{F}$, $g(\boldsymbol{x}) - \alpha_0 - \boldsymbol{b}^T\boldsymbol{x}$ belongs to $\mathcal{F}$. Since $\hat{f}_n$ is a minimizer of (2), we have
\[\sum_{i = 1}^n (y_i - \hat{f}_n(\boldsymbol{x}_i)^2 \leq  \sum_{i = 1}^n (y_i -(g(\boldsymbol{x}_i) - \alpha_0 - \boldsymbol{b}^T \boldsymbol{x}_i))^2,
\]
or equivalently,
\[\sum_{i = 1}^n (y_i + \alpha_0 + \boldsymbol{b}^T \boldsymbol{x}_i - (\hat{f}_n(\boldsymbol{x}_i) +  \alpha_0 + \boldsymbol{b}^T \boldsymbol{x}_i ))^2 \leq  \sum_{i = 1}^n (y_i + \alpha_0 + \boldsymbol{b}^T \boldsymbol{x}_i - g(\boldsymbol{x}_i))^2.
\]
So,
\[\sum_{i = 1}^n (y_i + \alpha_0 + \boldsymbol{b}^T \boldsymbol{x}_i - \hat{g}_n(\boldsymbol{x}_i))^2 \leq \sum_{i = 1}^n (y_i + \alpha_0 + \boldsymbol{b}^T \boldsymbol{x}_i - g(\boldsymbol{x}_i))^2\]
 for any $g \in \mathcal{F}$, and hence, $\hat{g}_n$ minimizes (\ref{eqn:01}).
\end{proof}

\subsubsection{Proof of Lemma 2}

\begin{proof}
We choose $\boldsymbol{x}_{n, 1}$ as follows. For any $c > 0$, consider the following hyperplane in $[0, 1]^d$:
\[H_c \triangleq \left\{ (x(1), \cdots, x(d)) \in [0, 1]^d : x(1) + \cdots + x(d) \leq c \right\}.\]
We let $c_0$ be the smallest positive real value of $c$ such that $H_{c}$ intersects with $\{\boldsymbol{x}_1, \cdots, \boldsymbol{x}_n\}$ at only one point. This is possible due to the fact that the $\boldsymbol{x}_i$'s have a continuous density function by \textbf{A3(ii)}. Let $\boldsymbol{x}_{n, 1}$ be the point that intersects with $H_{c_0}$; see Figure \ref{figure:01} for an illustration of  $H_{c_0}$ and $\boldsymbol{x}_{n, 1}$ when $d = 2$ and $n = 7$.  Mathematically, $\boldsymbol{x}_{n, 1} = \boldsymbol{x}_k$, where $k = \argmin_{i \in \{1, \cdots, n\}} x_i(1) + \cdots + x_i(d)$. By \textbf{A1}, $\boldsymbol{x}_{n, 1}$ exists uniquely almost surely.
\begin{figure}
\centering
\caption{An illustration of $H_{c_0}$ and $\boldsymbol{x}_{n, 1}$ when $d = 2$ and $n = 7$.}
\includegraphics[scale = 0.4]{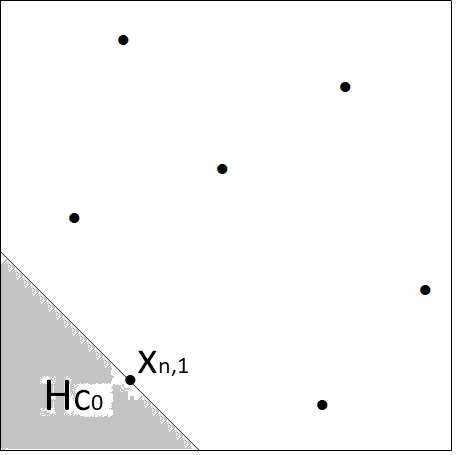}
\label{figure:01}
\end{figure}
Since the $\boldsymbol{x}_i$'s are distinct a.s., there exists $\delta > 0$ such that the rest of the $\boldsymbol{x}_i$'s belong to  \[\left\{ (x(1), \cdots, x(d)) \in [0, 1]^d : x(1) + \cdots + x(d) > c  + \delta\right\}.\]
It should be noted that there exists a convex function $h$ such that $h(\boldsymbol{x}_{n, 1}) = 1$ and $h(\boldsymbol{x}_j) = 0$ for all $\boldsymbol{x}_j$'s satisfying $\boldsymbol{x}_j \neq \boldsymbol{x}_{n, 1}$.

By (2.10) in Lemma 2.1 of \cite{Groeneboom1996} (or Lemma 2.4 (i) and (ii) of \citealp{seijo2011nonparametric}), $\sum_{i = 1}^n h(\boldsymbol{x}_i)(\hat{f}_n(\boldsymbol{x}_i) - y_i) \geq 0$, or equivalently, $\hat{f}_n(\boldsymbol{x}_{n, 1}) - y_{n, 1} \geq 0$, where $y_{n, 1}$ is the observed value at $\boldsymbol{x}_{n, 1}$. So,
\begin{equation}
\label{eqn:02} \hat{f}_n (\boldsymbol{x}_{n, 1}) \geq y_{n, 1}.
\end{equation}
Note that since $f_0$ is convex and differentiable, it is continuously differentiable by Theorem 25.5 of \cite{Rockafella1997}. Also, since $\nabla f_0 (\boldsymbol{0})< \boldsymbol{0}$, there exists a neighborhood of $\boldsymbol{0}$, where $\nabla f_0(\boldsymbol{x}) < \boldsymbol{0}$ for all $\boldsymbol{x}$ in that neighborhood. By \textbf{A1}, \textbf{A2}, \textbf{A3(i)}, and \textbf{A4}--\textbf{A6}, $\boldsymbol{\beta}_n \in \partial \hat{f}_n (\boldsymbol{x})$ converges to $\nabla f_0 (\boldsymbol{x})$ a.s.\ as $n \rightarrow \infty$ uniformly over $[\delta, 1- \delta]^d$ for any $\delta > 0$ (See Theorem 3.1 (iv) of \citealp{seijo2011nonparametric}). By the convexity of $\hat{f}_n$,  we can conclude that there exists a neighborhood of $\boldsymbol{0}$ such that $\boldsymbol{\beta} < \boldsymbol{0}$ for $\boldsymbol{\beta} \in \partial \hat{f}_n (\boldsymbol{x}) < \boldsymbol{0}$ for all $\boldsymbol{x}$ in that neighborhood and for $n$ sufficiently large. Since $\boldsymbol{x}_{n, 1} \rightarrow \boldsymbol{0}$ as $n \rightarrow \infty$ a.s., we can conclude
\begin{equation}
\label{eqn:03}\hat{f}_n(\boldsymbol{0}) > \hat{f}_n(\boldsymbol{x}_{n, 1})
\end{equation}
for $n$ sufficiently large a.s. By (\ref{eqn:02}) and (\ref{eqn:03}), $\hat{f}_n (\boldsymbol{0}) > y_{n, 1}$, and hence,
\begin{eqnarray}
\hat{f}_n (\boldsymbol{0}) - f_0 (\boldsymbol{0}) & > & y_{n, 1} - f_0(\boldsymbol{x}_{n, 1}) + f_0 (\boldsymbol{x}_{n, 1}) - f_0 (\boldsymbol{0})\nonumber\\
\label{eqn:04}& \geq & \varepsilon_1 + o(1)
\end{eqnarray}
for $n$ sufficiently large a.s.

Since $\varepsilon_1$ is a mean zero non-degenerate random variable, we have
\begin{equation}
\label{eqn:05}\mathbb{P}(\varepsilon_1 > \epsilon_0) > 0
\end{equation}
for some $\epsilon_0>0$.

By (\ref{eqn:04}) and (\ref{eqn:05}),
\[\liminf_{n \rightarrow \infty} \mathbb{P} (\hat{f}_n(\boldsymbol{0}) > f_0 (\boldsymbol{0}) + \epsilon_0) > 0,\]
which completes the proof of Lemma 2.
\end{proof}

\subsubsection{Proof of Theorem 1}
\begin{proof} Suppose $\nabla f_0(\boldsymbol{0}) < \boldsymbol{0}$. Then Theorem \ref{theorem: inconsistency} is followed by  Lemma \ref{lemma:02}. Suppose $\partial f_0 (\boldsymbol{0}) < \boldsymbol{0}$ does not hold. Then, there exists $\boldsymbol{b} \in \mathbb{R}^d$ such that $\nabla [f_0(\boldsymbol{x}) - \boldsymbol{b} \boldsymbol{x}] |_{\boldsymbol{x} = \boldsymbol{0}}  < \boldsymbol{0}$. Let $\hat{h}_n$ be the minimizer of
\begin{equation}
\label{eqn:06} \sum_{i = 1}^n (y_i - \boldsymbol{b}^T\boldsymbol{x}_i - f(\boldsymbol{x}_i))^2
\end{equation}
over $f \in \mathcal{F}$. Then, by Lemma \ref{lemma:01}, $\hat{f}_n (\boldsymbol{x}) - \boldsymbol{b}^T \boldsymbol{x}_i$ becomes a solution to (\ref{eqn:06}). Since (\ref{eqn:06}) can be written as $\sum_{i = 1}^n (f_0(\boldsymbol{x}) - \boldsymbol{b}^T\boldsymbol{x}_i  + \varepsilon_i - f(\boldsymbol{x}_i))^2$ and  $f_0(\boldsymbol{x}) - \boldsymbol{b}^T\boldsymbol{x}$ has a subgradient at $\boldsymbol{0}$ which is less than $\boldsymbol{0}$, Lemma \ref{lemma:02} implies there exists $\epsilon_0 > 0$ such that for any $\epsilon < \epsilon_0$,
\[\liminf_{n \rightarrow \infty} \mathbb{P}(\hat{f}_n (\boldsymbol{0}) + \boldsymbol{b}^T\boldsymbol{0} > f_0(\boldsymbol{0}) + \boldsymbol{b}^T\boldsymbol{0}  + \epsilon) > 0.\]
This proves Theorem \ref{theorem: inconsistency} when $\partial f_0(\boldsymbol{0}) < 0$ doesn't hold.
\end{proof}

\subsubsection{Proof of Theorem 2}
\begin{proof} We will prove that Theorem \ref{theorem: unboundedness} follows from Theorem \ref{theorem: inconsistency}. Let $\epsilon_0$ be given as in Lemma \ref{lemma:02}. Let $M > 0$ be given. Since $\hat{f}_n$ is convex, for any $\boldsymbol{x} \in [0, 1]^d \backslash \{\boldsymbol{0}\}$ and any $\boldsymbol{\beta} \in \partial \hat{f}_n (\boldsymbol{0})$, we have $\hat{f}_n (\boldsymbol{\boldsymbol{0}}) + \boldsymbol{\beta}^T \boldsymbol{x} \leq \hat{f}_n(\boldsymbol{x})$, and hence,
\[\hat{f}_n (\boldsymbol{0}) \leq \hat{f}_n (\boldsymbol{x}) + \|\boldsymbol{\beta}\| \|\boldsymbol{x}\|.\]
Since
\[\mathbb{P} (\hat{f}_n (\boldsymbol{0}) > f_0 (\boldsymbol{0}) + \epsilon_0) \leq \mathbb{P}(\hat{f}_n (\boldsymbol{x}) > f_0 (\boldsymbol{0}) + \epsilon_0/2) + \mathbb{P} \left(\min_{\boldsymbol{\beta} \in \partial \hat{f}_n(\boldsymbol{0})} \|\boldsymbol{\beta}\| \|\boldsymbol{x}\| > \epsilon_0/2\right),\]
taking the liminf on both sides yields
\begin{eqnarray}
\lefteqn{\liminf_{n \rightarrow \infty} \mathbb{P} (\hat{f}_n (\boldsymbol{0}) > f_0 (\boldsymbol{0}) + \epsilon_0)}\nonumber\\
\label{eqn:07}& \leq & \liminf_{n \rightarrow \infty}\left[\mathbb{P}(\hat{f}_n (\boldsymbol{x}) > f_0 (\boldsymbol{0}) + \epsilon_0/2) + \mathbb{P} \left(\min_{\boldsymbol{\beta} \in \partial \hat{f}_n(\boldsymbol{0})} \|\boldsymbol{\beta}\| > \epsilon_0/(2\|\boldsymbol{x}\|)\right)\right].
\end{eqnarray}
Since $\hat{f}_n$ converges to $f_0$ uniformly on $[\delta, 1 - \delta]^d$ a.s.\ for any $\delta > 0$ (Theorem 3.1 of \citealp{seijo2011nonparametric}) and $f_0$ is continuous (\textbf{A6}), we can take $\boldsymbol{x}$ close enough to $\boldsymbol{0}$ so that $\lim_{n \rightarrow \infty}\mathbb{P}(\hat{f}_n (\boldsymbol{x}) > f_0 (\boldsymbol{0}) + \epsilon_0/2)= 0$. From (\ref{eqn:07}), it follows that
\begin{equation}
\label{eqn:39}\liminf_{n \rightarrow \infty} \mathbb{P} (\hat{f}_n (\boldsymbol{0}) > f_0 (\boldsymbol{0}) + \epsilon_0)
 \leq \liminf_{n \rightarrow \infty}\mathbb{P} \left(\min_{\boldsymbol{\beta} \in \partial \hat{f}_n(\boldsymbol{0})} \|\boldsymbol{\beta}\| > \epsilon_0/(2\|\boldsymbol{x}\|)\right).
\end{equation}
We can also take $\boldsymbol{x}$ close enough to $\boldsymbol{0}$ so that $\epsilon_0/(2\|\boldsymbol{x}\|) > M$. From (\ref{eqn:39}) and Theorem \ref{theorem: inconsistency}, we obtain
\[\liminf_{n \rightarrow \infty} \mathbb{P} \left(\sup_{\boldsymbol{\beta} \in \partial \hat{f}_n(\boldsymbol{0})} \|\boldsymbol{\beta}\| > M\right) > \epsilon_1,\]
for some real number $\epsilon_1 > 0$, which completes the proof of Theorem \ref{theorem: unboundedness}.
\end{proof}

\subsubsection{Proof of Theorem 3}
We first prove (i), and then show that (i) implies (ii). 
\begin{proof}
The proof of (i) consists of 13 steps.

\underline{Step 1:} We use \textbf{A5}, \textbf{A7}, and the fact that $\hat{f}_{n, A}$ is a minimizer of $\sum_{i = 1}^n (y_i - f(\boldsymbol{x}_i))^2$ over $f \in \mathcal{F}_A$ to obtain
\begin{equation}
\label{eqn:08}
\frac{1}{n}\sum_{i = 1}^n (\hat{f}_{n, A}(\boldsymbol{x}_i) - f_0(\boldsymbol{x}_i))^2 \leq \frac{2}{n} \sum_{i = 1}^n \varepsilon_i (\hat{f}_{n, A}(\boldsymbol{x}_i) - f_0(\boldsymbol{x}_i)).
\end{equation}
To prove (\ref{eqn:08}), note that $f_0 \in \mathcal{F}_A$ by \textbf{A5} and \textbf{A7} and that $\hat{f}_{n, A}$ is a minimizer of (14). So, we have
\[\frac{1}{n} \sum_{i = 1}^n (y_i - \hat{f}_{n, A} (\boldsymbol{x}_i))^2 \leq \frac{1}{n} \sum_{i = 1}^n (y_i - f_0 (\boldsymbol{x}_i))^2, \]
or equivalently,
\[\frac{1}{n} \sum_{i = 1}^n (f_0(\boldsymbol{x}_i) + \varepsilon_i - \hat{f}_{n, A} (\boldsymbol{x}_i))^2 \leq \frac{1}{n} \sum_{i = 1}^n (f_0(\boldsymbol{x}_i) + \varepsilon_i - f_0 (\boldsymbol{x}_i))^2, \]
which leads to (\ref{eqn:08}).

\underline{Step 2:} We use \textbf{A1}, \textbf{A2}, \textbf{A4}, and (\ref{eqn:08}) to prove that there exists a constant $c_0$ such that
\begin{equation}
\label{eqn:09}\frac{1}{n}\sum_{i = 1}^n \hat{f}_{n, A}(\boldsymbol{x}_i)^2 \leq c_0
\end{equation}
for $n$ sufficiently large a.s.

To prove (\ref{eqn:09}), note that
\begin{eqnarray}
\frac{1}{n}\sum_{i = 1}^n \hat{f}_{n, A}(\boldsymbol{x}_i)^2 & \leq & \frac{1}{n}\sum_{i = 1}^n (\hat{f}_{n, A}(\boldsymbol{x}_i) - f_0 (\boldsymbol{x}_i) + f_0 (\boldsymbol{x}_i))^2\nonumber\\
\label{eqn:10}& \leq & \frac{2}{n} \sum_{i = 1}^n (\hat{f}_{n, A}(\boldsymbol{x}_i) - f_0 (\boldsymbol{x}_i))^2 + \frac{2}{n}\sum_{i = 1}^n f_0(\boldsymbol{x}_i)^2.
\end{eqnarray}
On the other hand, applying the Cauchy-Schwarz inequality to the right-hand side of (\ref{eqn:08}) yields
\[\frac{1}{n}\sum_{i = 1}^n (\hat{f}_{n, A}(\boldsymbol{x}_i) - f_0 (\boldsymbol{x}_i))^2\leq 2 \sqrt{\frac{1}{n}\sum_{i = 1}^n \varepsilon_i^2}\sqrt{\frac{1}{n}\sum_{i = 1}^n (\hat{f}_{n, A} (\boldsymbol{x}_i) - f_0 (\boldsymbol{x}_i))^2},\]
so we have 
\begin{equation}
\label{eqn:11} \frac{1}{n}\sum_{i = 1}^n (\hat{f}_{n, A}(\boldsymbol{x}_i) - f_0 (\boldsymbol{x}_i))^2 \leq \frac{4}{n} \sum_{i = 1}^n \varepsilon_i^2.
\end{equation}
Combining (\ref{eqn:10}) and (\ref{eqn:11}) yields
\[\frac{1}{n} \sum_{i = 1}^n \hat{f}_{n, A} (\boldsymbol{x}_i)^2 \leq \frac{8}{n} \sum_{i = 1}^n \varepsilon_i^2 + \frac{2}{n}\sum_{i = 1}^n f_0 (\boldsymbol{x}_i)^2.\]
By \textbf{A1}, \textbf{A2}, \textbf{A4}, and the strong law of large numbers,
\[\frac{1}{n}\sum_{i = 1}^n \hat{f}_{n, A}(\boldsymbol{x}_i)^2 \leq 8 \mathbb{E}[\varepsilon_1^2] + 2 \mathbb{E}[f_0(\boldsymbol{x}_1)^2] + 1 \triangleq c_0\]
for $n$ sufficiently large a.s.

\underline{Step 3:} In the following steps, we will use the following lemma, which is Lemma 1 of \citet{LuoLim2016}.

Let $\boldsymbol{e}_0 = (0, 0, \cdots, 0)^T$. For $1\leq i\leq d$, let $\boldsymbol{e}_i \in \mathbb{R}^d$ be the vector of zeros except 1 in the $i$th entry. Let $\boldsymbol{v}_* = (1/(4d), 1/d, \cdots,$ $1/d)$.
Let $A_i$ be defined as follows:
\begin{eqnarray*}
A_0 & = & \left\{\boldsymbol{x} \in [0, 1]^d : \|\boldsymbol{x} - \boldsymbol{e}_0\| \leq \tau \right\},\\
A_1 & = & [1/2, 1]\times [0, 1] \times \dots \times [0, 1] \subset [0, 1]^d,\\
A_i & = & \left\{\boldsymbol{x} \in [0, 1]^d : \|\boldsymbol{x} - \boldsymbol{e}_i\| \leq \tau \right\} \mbox{ for } 2 \leq i \leq d,\\
A_{d + 1} & = & \left\{\boldsymbol{x} \in [0, 1]^d : \|\boldsymbol{x} - \boldsymbol{v}_*\| \leq \tau \right\}.
\end{eqnarray*}
Then there exists a positive constant $\tau_*$ such that $0 \leq \tau \leq \tau_*$ implies for any $\boldsymbol{y}$ in $A_{d + 1}$ and $\boldsymbol{z}_i$ in $A_i$  for $0 \leq i\leq d$,
there exist nonnegative real numbers $p_0, p_1, \dots, p_d$ summing to one such that
\[p_0 \boldsymbol{z}_0 + p_1 \boldsymbol{z}_1 + \dots + p_d \boldsymbol{z}_d = \boldsymbol{y}\]
and that $p_1 \geq 1/(16d)$.

\underline{Step 4:} In the following steps, we will also use the following lemma, which is Lemma 2 of \citet{LuoLim2016}.

For $0 \leq i \leq d$, let $\boldsymbol{u}_i$ be the vector identical to $\boldsymbol{e}_i$ except that its first element is one minus $\boldsymbol{e}_i$'s first element.
Let $\boldsymbol{w}_* = (1-1/(4d), 1/d, \cdots, 1/d)$.
Let $B_i$ be defined as follows:
\begin{eqnarray*}
B_0 & = & \left\{\boldsymbol{x} \in [0, 1]^d : \|\boldsymbol{x} - \boldsymbol{u}_0\| \leq \tau \right\},\\
B_1 & = & [0, 1/2]\times [0, 1] \times \dots \times [0, 1] \subset [0, 1]^d,\\
B_i & = & \left\{\boldsymbol{x} \in [0, 1]^d : \|\boldsymbol{x} - \boldsymbol{u}_i\| \leq \tau \right\} \mbox{ for } 2 \leq i \leq d,  \\
B_{d + 1} & = & \left\{\boldsymbol{x} \in [0, 1]^d : \|\boldsymbol{x} - \boldsymbol{w}_*\| \leq \tau \right\}.
\end{eqnarray*}
Then, there exists a positive constant $\tau_*$ such that $0 \leq \tau \leq \tau_*$ implies for any $\boldsymbol{y}$ in $B_{d + 1}$ and $\boldsymbol{z}_i$ in $B_i$  for $0 \leq i \leq d$,
there exist nonnegative real numbers $p_0, p_1, \dots, p_d$ summing to one such that
\[p_0 \boldsymbol{z}_0 + p_1 \boldsymbol{z}_1 + \dots + p_d \boldsymbol{z}_d = \boldsymbol{y}\]
and that $p_1 \geq 1/(16d)$.

\underline{Step 5:} We use Step 2, Step 3, Step 4, and \textbf{A3(i)} to establish that there exists a constant $r_0$ such that 
\begin{equation}
\label{eqn:25}
\inf_{\boldsymbol{x}_i \in A_j} |\hat{f}_{n, A} (\boldsymbol{x}_i)| \leq r_0
\end{equation} 
and
\begin{equation}
\label{eqn:26}
\inf_{\boldsymbol{x}_i \in B_j} |\hat{f}_{n, A} (\boldsymbol{x}_i)| \leq r_0
\end{equation} 
for all $j = 0, 1, \cdots, d + 1$ and for $n$ sufficiently large a.s. 

To fill in the details, we note that, for any $j \in \{0, 1, \cdots, d + 1\}$ and $r > 0$,
\begin{equation}
\frac{1}{n}\sum_{i = 1}^n I (\boldsymbol{x}_i \in A_j, |\hat{f}_{n, A} (\boldsymbol{x}_i)| \leq r) = \frac{1}{n}\sum_{i = 1}^n I(\boldsymbol{x}_i \in A_j) - \frac{1}{n}\sum_{i = 1}^n I(\boldsymbol{x}_i \in A_j, |\hat{f}_{n, A} (\boldsymbol{x}_i)| > r).
\end{equation}
By Step 2 and Markov inequality,
\[\frac{1}{n} \sum_{i = 1}^n I (\boldsymbol{x}_i \in A_j, |\hat{f}_{n, A} (\boldsymbol{x}_i)| > r) \leq \frac{1}{r^2} \cdot \frac{1}{n}\sum_{i = 1}^n \hat{f}_{n, A} (\boldsymbol{x}_i)^2\]
for $n$ sufficiently large. Choose $r_0$ so large that $(c_0 + 1)/r_0^2 \leq 1/2 \min_{0 \leq j \leq d + 1} \mathbb{P}(\boldsymbol{x}_1 \in A_j)$, then 
\[\frac{1}{n} \sum_{i = 1}^n I (\boldsymbol{x}_i \in A_j, |\hat{f}_{n, A} (\boldsymbol{x}_i)| \leq r_0) > \frac{1}{2} \min_{0 \leq j \leq d + 1} \mathbb{P}(\boldsymbol{x}_1 \in A_j)\]
for $n$ sufficiently large a.s. By \textbf{A3(i)}, $ \min_{0 \leq j \leq d + 1} \mathbb{P}(\boldsymbol{x}_1 \in A_j) > 0$. By the strong law of large numbers, 
\[\frac{1}{n}\sum_{i=1}^n I(\boldsymbol{x}_i \in A_j) \rightarrow \mathbb{P}(\boldsymbol{x}_1 \in A_j)\]
as $n \rightarrow \infty$ a.s.\ and
\[\liminf_{n \rightarrow \infty} \frac{1}{n}\sum_{i=1}^n I(\boldsymbol{x}_i \in A_j, |\hat{f}_{n, A}(\boldsymbol{x}_i)|\leq r) >0.\]
So, for $n$ sufficiently large, there exist $\boldsymbol{x}_{I(j)} \in A_j$ with $1 \leq I(j) \leq n$ such that $|\hat{f}_{n, A} (\boldsymbol{x}_{I(j)})| \leq r_0$. So, (\ref{eqn:25}) is established. (\ref{eqn:26}) follows in a similar way.

\underline{Step 6:} We use Step 3, Step 5, and the convexity of $\hat{f}_{n, A}$ to establish that 
\[\inf_{\boldsymbol{x} \in [0, 1]^d} \hat{f}_{n, A} (\boldsymbol{x}) \geq -64 r_0 d\]
for $n$ sufficiently large a.s.

We will first establish $\inf_{\boldsymbol{x} \in A_1} \hat{f}_{n, A} (\boldsymbol{x}) \geq -64 r_0 d$ for $n$ sufficiently large a.s., and then show $\inf_{\boldsymbol{x} \in B_1} \hat{f}_{n, A} (\boldsymbol{x}) \geq -64r_0 d$ for $n$ sufficiently large a.s.

To fill in the details, suppose that there exists $\boldsymbol{z}_1 \in A_1$ and 
\begin{equation*}
\hat{f}_{n, A} (\boldsymbol{z}_1) < -64 r_0 d
\end{equation*}
for some $n$ sufficiently large.

By Step 3 and Step 5, for any $\boldsymbol{y} \in A_{d + 1}$, there exist  $\boldsymbol{z}_0 \in A_0, \boldsymbol{z}_2 \in A_2, \cdots, \boldsymbol{z}_d \in A_d$ and nonnegative real numbers $p_0, p_1, \cdots, p_d$ satisfying
\begin{align}
\label{eqn:28}&|\hat{f}_{n, A} (\boldsymbol{z}_i)| \leq r_0 \mbox{ for } i = 0, 2, \cdots, d,\\
&p_0 \boldsymbol{z}_0 + p_1 \boldsymbol{z}_1 + p_2 \boldsymbol{z}_2 + \cdots + p_d \boldsymbol{z}_d = \boldsymbol{y},\nonumber\\
&p_0 + p_1  + \cdots + p_d = 1, \mbox{ and}\nonumber\\
&p_1 > 1/(16d).\nonumber
\end{align}

By the convexity of $\hat{f}_{n, A}$, we have 
\begin{eqnarray*}
\hat{f}_{n, A} (\boldsymbol{y}) & \leq & \hat{f}_{n, A} (p_0 \boldsymbol{z}_0 + p_1 \boldsymbol{z}_1 + \cdots + p_d \boldsymbol{z}_d)\\
& \leq & p_0 \hat{f}_{n, A} (\boldsymbol{z}_0) + p_1 \hat{f}_{n, A} (\boldsymbol{z}_1) + p_2 \hat{f}_{n, A} (\boldsymbol{z}_2) + \cdots + p_d \hat{f}_{n, A} (\boldsymbol{z}_d).
\end{eqnarray*}
Since
$\hat{f}_{n, A} (\boldsymbol{z}_1) \leq -64 r_0 d$, $p_1 > 1/(16d)$, and (\ref{eqn:28}) imply
\[\hat{f}_{n, A} (\boldsymbol{y}) \leq -3r_0,\]
and hence, 
\[\inf_{\boldsymbol{x}\in A_{d+1}}|\hat{f}_{n, A}(\boldsymbol{x})| > r_0,\]
which contradicts Step 5. So, we have $\inf_{\boldsymbol{x} \in A_1} \hat{f}_{n, A} (\boldsymbol{x}) \geq -64r_0 d$ for all $n$ sufficiently large. Similarly, we have $\inf_{\boldsymbol{x} \in B_1} \hat{f}_{n, A} (\boldsymbol{x}) \geq -64r_0 d$.

\underline{Step 7:} We use the convexity of $\hat{f}_{n, A}$ and Step 2 to show that for any $\delta > 0$, there exists a constant $c_{\delta}$ satisfying 
\[\sup_{\boldsymbol{x} \in [\delta, 1 - \delta]^d} \hat{f}_{n, A} (\boldsymbol{x}) \leq c_{\delta}\]
for $n$ sufficiently large a.s. For a complete proof, see the proof of Lemma 4 of \cite{LuoLim2016}.

\underline{Step 8:} There exists a constant $c_1$ such that
\begin{equation}
\label{eqn:13} |\hat{f}_{n, A} (\boldsymbol{x}) - \hat{f}_{n, A}(\boldsymbol{y})| \leq c_1 \|\boldsymbol{x} - \boldsymbol{y}\|
\end{equation}
for any $\boldsymbol{x}, \boldsymbol{y} \in [0, 1]^d$ and $n = 1, 2, \cdots$ a.s.

(\ref{eqn:13}) follows from the fact that $\hat{f}_{n, A}$ is a convex function having its subgradients bounded uniformly over $[0, 1]^d$ and $n$, so $\hat{f}_{n, A}$ is Lipshitz with a Lipshitz constant uniformly on $n$ and over $[0, 1]^d$.

\underline{Step 9:} We use Step 6, Step 7, and Step 8 to establish there exists a constant $\gamma$ such that
\begin{equation}
\label{eqn:14}\sup_{\boldsymbol{x} \in [0, 1]^d} |\hat{f}_{n, A} (\boldsymbol{x})| \leq \gamma
\end{equation}
for $n$ sufficiently large a.s.

To show (\ref{eqn:14}), we note that by Step 6 and Step 7, there is a constant $c$ satisfying
\begin{equation}
\label{eqn:15} \sup_{\boldsymbol{x} \in [0.1, 0.9]^d} |\hat{f}_{n, A}(\boldsymbol{x})| < c
\end{equation}
for $n$ sufficiently large a.s. Let $\boldsymbol{x}_0$ be any arbitrary point in $[0.1, 0.9]^d$. For any $\boldsymbol{y} \in [0, 1]^d\backslash [0.1, 0.9]^d$, Step 8 and (\ref{eqn:15}) imply
\[|\hat{f}_{n, A} (\boldsymbol{y})| \leq |\hat{f}_{n, A} (\boldsymbol{x})| + c_1 \| \boldsymbol{x} - \boldsymbol{y}\| \leq c + c_1 d \triangleq \gamma\]
for $n$ sufficiently large a.s., which proves Step 9.

\underline{Step 10:} We note that for any $\epsilon> 0$, there is a finite number of functions $f_1, \cdots, f_r$ in 
\[\tilde{\mathcal{F}}_A \triangleq \{f:[0, 1]^d \rightarrow \mathbb{R}: f \mbox{ is convex, }| f (\boldsymbol{x}) - f(\boldsymbol{y})| \leq c_1 \|\boldsymbol{x} - \boldsymbol{y}\| \mbox{ for } \boldsymbol{x}, \boldsymbol{y} \in [0, 1]^d, \sup_{\boldsymbol{x} \in [0, 1]^d} |f(\boldsymbol{x})| \leq \gamma\}\] with $r = r(\epsilon)$ satisfying, for any $f \in \tilde{\mathcal{F}}_A$,
\begin{equation}
\label{eqn:16} \sup_{\boldsymbol{x} \in [0, 1]^d} |f(\boldsymbol{x}) - f_i (\boldsymbol{x})| < \epsilon
\end{equation}
for some $i \in \{1, \cdots, r\}$. (\ref{eqn:16}) follows from Step 8, Step 9, and Theorem 6 of \cite{Bronshtein1976}.

\underline{Step 11:} We now use Step 10 and the strong law of large numbers to show that
\begin{equation}
\label{eqn:17} \limsup_{n \rightarrow \infty} \frac{1}{n}\sum_{i = 1}^n \varepsilon_i (\hat{f}_{n, A}(\boldsymbol{x}_i) - f_0 (\boldsymbol{x}_i)) \leq 0
\end{equation}
a.s. To fill in the details, let $\epsilon > 0$ be given and select $f_1, \cdots, f_r$ as suggested in Step 10. Then, for each $j \in \{1, \cdots, r\}$, we have
\begin{eqnarray*}
\lefteqn{\frac{1}{n} \sum_{i = 1}^n \varepsilon_i (\hat{f}_{n, A} (\boldsymbol{x}_i) - f_0 (\boldsymbol{x}_i))}\\
& \leq & \frac{1}{n} \sum_{i = 1}^n \varepsilon_i (\hat{f}_{n, A} (\boldsymbol{x}_i) - f_j (\boldsymbol{x}_i)) +  \frac{1}{n} \sum_{i = 1}^n \varepsilon_i (f_j(\boldsymbol{x}_i) - f_0 (\boldsymbol{x}_i))\\
& \leq & \frac{1}{n} \sum_{i = 1}^n |\varepsilon_i| \sup_{\boldsymbol{x} \in [0, 1]^d} |\hat{f}_{n, A} (\boldsymbol{x}) - f_j (\boldsymbol{x})| +  \max_{1 \leq j \leq r}\frac{1}{n} \sum_{i = 1}^n \varepsilon_i (f_j(\boldsymbol{x}_i) - f_0 (\boldsymbol{x}_i))
\end{eqnarray*}

So,
\begin{eqnarray*}
\lefteqn{\frac{1}{n}\sum_{i = 1}^n \varepsilon_i (\hat{f}_{n, A} (\boldsymbol{x}_i) - f_0(\boldsymbol{x}_i))}\\
& \leq & \left[\min_{1 \leq j \leq r} \sup_{\boldsymbol{x} \in [0, 1]^d}  |\hat{f}_{n, A} (\boldsymbol{x}) - f_j (\boldsymbol{x})|\right]  \left[  \frac{1}{n} \sum_{i = 1}^n |\varepsilon_i|\right] +  \max_{1 \leq j \leq r}\frac{1}{n} \sum_{i = 1}^n \varepsilon_i (f_j(\boldsymbol{x}_i) - f_0 (\boldsymbol{x}_i))\\
& \leq & \frac{\epsilon}{n}\sum_{i = 1}^n |\varepsilon_i|  + \max_{1 \leq j \leq r}\frac{1}{n} \sum_{i = 1}^n \varepsilon_i (f_j(\boldsymbol{x}_i) - f_0 (\boldsymbol{x}_i))\\
& \leq & \epsilon(\mathbb{E}[|\varepsilon_1|] + 1) + \epsilon
\end{eqnarray*}
for $n$ sufficiently large a.s.\ by \textbf{A1} and \textbf{A2}, and hence, (\ref{eqn:17}) follows.

\underline{Step 12:} By combining Step 1 and Step 11, we obtain
\[\frac{1}{n}\sum_{i = 1}^n (\hat{f}_{n, A} (\boldsymbol{x}_i) - f_0 (\boldsymbol{x}_i))^2 \rightarrow 0\]
as $n \rightarrow \infty$ a.s.

\underline{Step 13:} We use Step 12 and the fact that $\hat{f}_{n, A}$ and $f_0$ are Lipschitz over $[0, 1]^d$ to complete the proof of (i).

Let $\epsilon > 0$ be given. Since $\Omega = [0, 1]^d$ is compact, there exists a finite collection of sets $S_1, \cdots, S_l$ covering $\Omega$, each having a diameter less than $\epsilon$, i.e., $\|\boldsymbol{x} - \boldsymbol{y}\| \leq \epsilon$ for any $\boldsymbol{x}, \boldsymbol{y} \in S_j$ and $1 \leq j \leq l$. By \textbf{A7} and Step 4,  there exists a Lipschitz constant, say $\lambda$, for both $\hat{f}_{n, A}$ and $f_0$ over $[0, 1]^d$. For each $\boldsymbol{x} \in S_j$ and $\boldsymbol{x}_i \in S_j$,
\begin{eqnarray*}
\lefteqn{|\hat{f}_{n, A}(\boldsymbol{x})- f_0 (\boldsymbol{x})|}\\
& \leq & |\hat{f}_{n, A}(\boldsymbol{x}) - \hat{f}_{n, A} (\boldsymbol{x}_i)| + |\hat{f}_{n, A} (\boldsymbol{x}_i) - f_0 (\boldsymbol{x}_i)| + |f_0 (\boldsymbol{x}_i) - f_0 (\boldsymbol{x})|\\
& \leq & 2 \epsilon \lambda + |\hat{f}_{n, A} (\boldsymbol{x}_i) - f_0 (\boldsymbol{x}_i)|.
\end{eqnarray*}
So, for any $\boldsymbol{x} \in S_j$,
\[|\hat{f}_{n, A} (\boldsymbol{x}) - f_0 (\boldsymbol{x})| \leq 2 \epsilon \lambda  + \min_{\boldsymbol{x}_i \in S_j} |\hat{f}_{n, A} (\boldsymbol{x}_i) - f_0 (\boldsymbol{x}_i)|.\]
Therefore,
\begin{eqnarray*}
\lefteqn{\sup_{\boldsymbol{x} \in S_j} |\hat{f}_{n, A}(\boldsymbol{x}) - f_0 (\boldsymbol{x})|}\\
& \leq & 2 \epsilon \lambda + \min_{\boldsymbol{x}_i\in S_j} |\hat{f}_{n, A} (\boldsymbol{x}_i) - f_0 (\boldsymbol{x}_i)|\\
& \leq & 2 \epsilon \lambda + \left( \frac{1}{n} \sum_{i = 1}^n |\hat{f}_{n, A} (\boldsymbol{x}_i) - f_0 (\boldsymbol{x}_i)|I (\boldsymbol{x}_i \in S_j)\right) \left(\frac{n}{\sum_{i = 1}^n I (\boldsymbol{x}_i \in S_j)} \right)\\
& \leq &  2 \epsilon \lambda + \sqrt{ \frac{1}{n} \sum_{i = 1}^n (\hat{f}_{n, A} (\boldsymbol{x}_i) - f_0 (\boldsymbol{x}_i))^2} \left(\frac{n}{\sum_{i = 1}^n I (\boldsymbol{x}_i \in S_j)} \right).
\end{eqnarray*}
By Step 12, \textbf{A1} and \textbf{A3(i)}, we conclude
\[\limsup_{n \rightarrow \infty} \sup_{\boldsymbol{x} \in S_j} |\hat{f}_{n, A} (\boldsymbol{x}) - f_0 (\boldsymbol{x})| \leq 2 \epsilon \lambda\]
a.s. Since $\epsilon$ is arbitrary and there are finitely many $S_j$'s, we conclude
\[\sup_{\boldsymbol{x} \in [0, 1]^d} |\hat{f}_{n, A} (\boldsymbol{x}) - f_0 (\boldsymbol{x})| \rightarrow 0\]
a.s.\ as $n \rightarrow \infty$, completing the proof of (i).

Next, we show that (i) implies (ii).

Suppose, on the contrary, that there exists $\epsilon_0 > 0$ and $\boldsymbol{z}_1, \boldsymbol{z}_2, \cdots \in [0, 1]^d$ such that
\[\|\boldsymbol{\beta} - \nabla f_0 (\boldsymbol{z}_n)\| \geq \epsilon_0\]
for some $\boldsymbol{\beta} \in \partial \hat{f}_{n, A} (\boldsymbol{z}_n)$ and infinitely many $n$ with a positive probability. It follows that there is $i \in \{1, \cdots, d\}$ satisfying
\begin{equation}
\label{eqn:18}|\boldsymbol{e}_i^T \boldsymbol{\beta} - \boldsymbol{e}_i^T \nabla f_0 (\boldsymbol{z}_n)| \geq \epsilon_0/d
\end{equation}
for some $\beta \in \partial \hat{f}_{n, A} (\boldsymbol{z}_n)$ and infinitely many $n$ with positive probability, where $\boldsymbol{e}_i \in \mathbb{R}^d$ is the vector of zeros except for 1 in the $i$th entry for $1 \leq i \leq d$.

(\ref{eqn:18}) implies either
\begin{equation}
\label{eqn:19}
\boldsymbol{e}_i^T \boldsymbol{\beta} \geq \boldsymbol{e}_i^T \nabla f_0 (\boldsymbol{z}_n) + \epsilon_0/d
\end{equation}
or
\begin{equation}
\label{eqn:20}
\boldsymbol{e}_i^T \nabla f_0 (\boldsymbol{z}_n) - \epsilon_0/d \geq \boldsymbol{e}_i^T \boldsymbol{\beta}
\end{equation}
for some $\boldsymbol{\beta} \in \partial \hat{f}_{n, A} (\boldsymbol{z}_n)$ and infinitely many $n$ with positive probability. We will consider the case where (\ref{eqn:19}) holds and reaches a contradiction. Similar arguments can be applied when (\ref{eqn:20}) holds.

When (\ref{eqn:19}) holds, there exists a subsequence $\boldsymbol{z}_{n_1}, \boldsymbol{z}_{n_2}, \cdots$ such that $\boldsymbol{z}_{n_k} \rightarrow \boldsymbol{z}_0 \in [0, 1]^d$ as $k \rightarrow \infty$. By the definition of the subgradient, for any $h > 0$,
\[\boldsymbol{e}_i^T\boldsymbol{\beta} \leq \frac{\hat{f}_{n_k, A} (\boldsymbol{z}_{n_k} + h \boldsymbol{e}_i) - \hat{f}_{n_k, A}(\boldsymbol{z}_{n_k})}{h}.\]
By (i) and the fact that $\hat{f}_{n_k, A}$ has its subgradients bounded uniformly on $n$, letting $k \rightarrow \infty$ yields
\begin{equation}
\label{eqn:21}\boldsymbol{e}_i^T \boldsymbol{\beta} \leq \frac{f_0 (\boldsymbol{z}_0  + h \boldsymbol{e}_i) - f_0 (\boldsymbol{z}_0)}{h}
\end{equation}
for any $h > 0$.

By (\ref{eqn:19}) and (\ref{eqn:21}), we have
\[\boldsymbol{e}_i^T \nabla f_0 (\boldsymbol{z}_{n_k}) + \epsilon_0 /d \leq \frac{f_0 (\boldsymbol{z}_0 + h \boldsymbol{e}_i) - f_0 (\boldsymbol{z}_0)}{h}\]
for $k \geq 1$.

By \textbf{A5} and \textbf{A6}, $f_0$ is continuously differentiable (Darboux theorem), so we have 
\begin{equation}
\label{eqn:22}\boldsymbol{e}_i^T \nabla f_0 (\boldsymbol{z}_0) + \epsilon_0 /d \leq \frac{f_0 (\boldsymbol{z}_0 + h \boldsymbol{e}_i) - f_0 (\boldsymbol{z}_0)}{h}\end{equation}
for any $h > 0$.

By letting $h \downarrow 0$ in (\ref{eqn:22}), we obtain
\[\boldsymbol{e}_i^T \nabla f_0(\boldsymbol{z}_0) + \epsilon_0 /d \leq \nabla f_0 (\boldsymbol{z}_0),\]
which is a contradiction.

Similarly, when we assume (\ref{eqn:20}) holds, we reach a contradiction. So, the proof of (ii) is completed.
\end{proof}

\subsubsection{Proof of Theorem 4}
\begin{proof} The proof of Theorem \ref{theorem: wrcr consistency} (i) is similar to that of Theorem \ref{theorem: alcr consistency} (i) with \textbf{A7}  replaced by \textbf{A8}.
The proof of Theorem \ref{theorem: wrcr consistency} (ii) is similar to that of Theorem \ref{theorem: alcr consistency} (ii) with \textbf{A7} replaced by \textbf{A8}
\end{proof}

\subsection{Additional experiments}
\setcounter{figure}{0}
\setcounter{table}{0}
\setcounter{equation}{0}

We consider the following true regression functions:
\begin{enumerate}
    \item[] Type I: \(f(\boldsymbol{x}) = \sum_{p=1}^d x_p^2\),
    \item[] Type II: \(f(\boldsymbol{x}) = \sum_{p=1}^d (x_p-0.2)^2\).
\end{enumerate}

\subsubsection{Estimation of convex functions}

We present some additional computational results on the estimation of convex functions. 
\begin{figure}[H]
    \centering
    \includegraphics[width=0.8\textwidth]{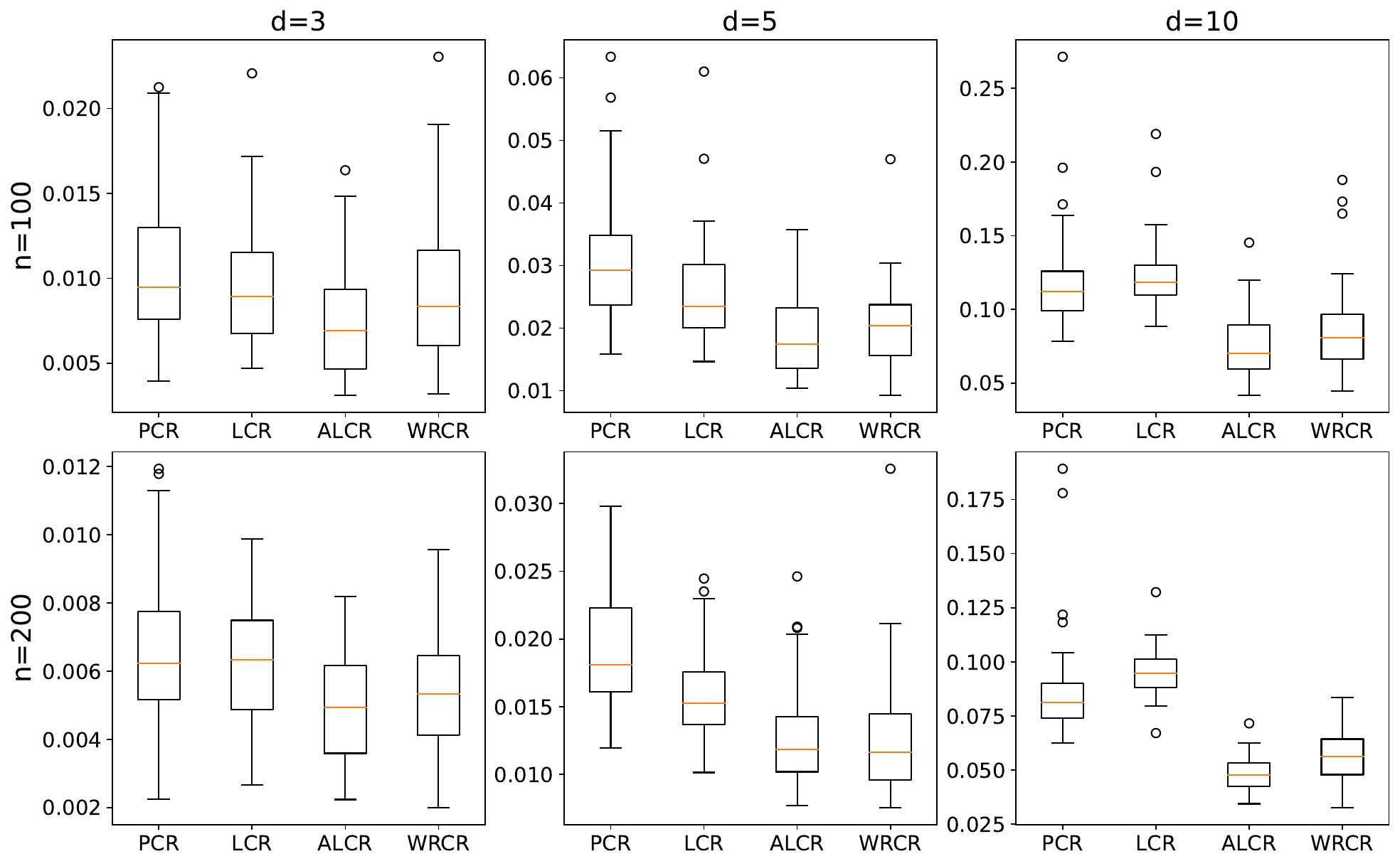}
    \caption{Table showing out-of-sample performance of four estimators for data Type I. }
    \label{fig: conva}
\end{figure}

\begin{figure}[H]
    \centering
    \includegraphics[width=0.8\textwidth]{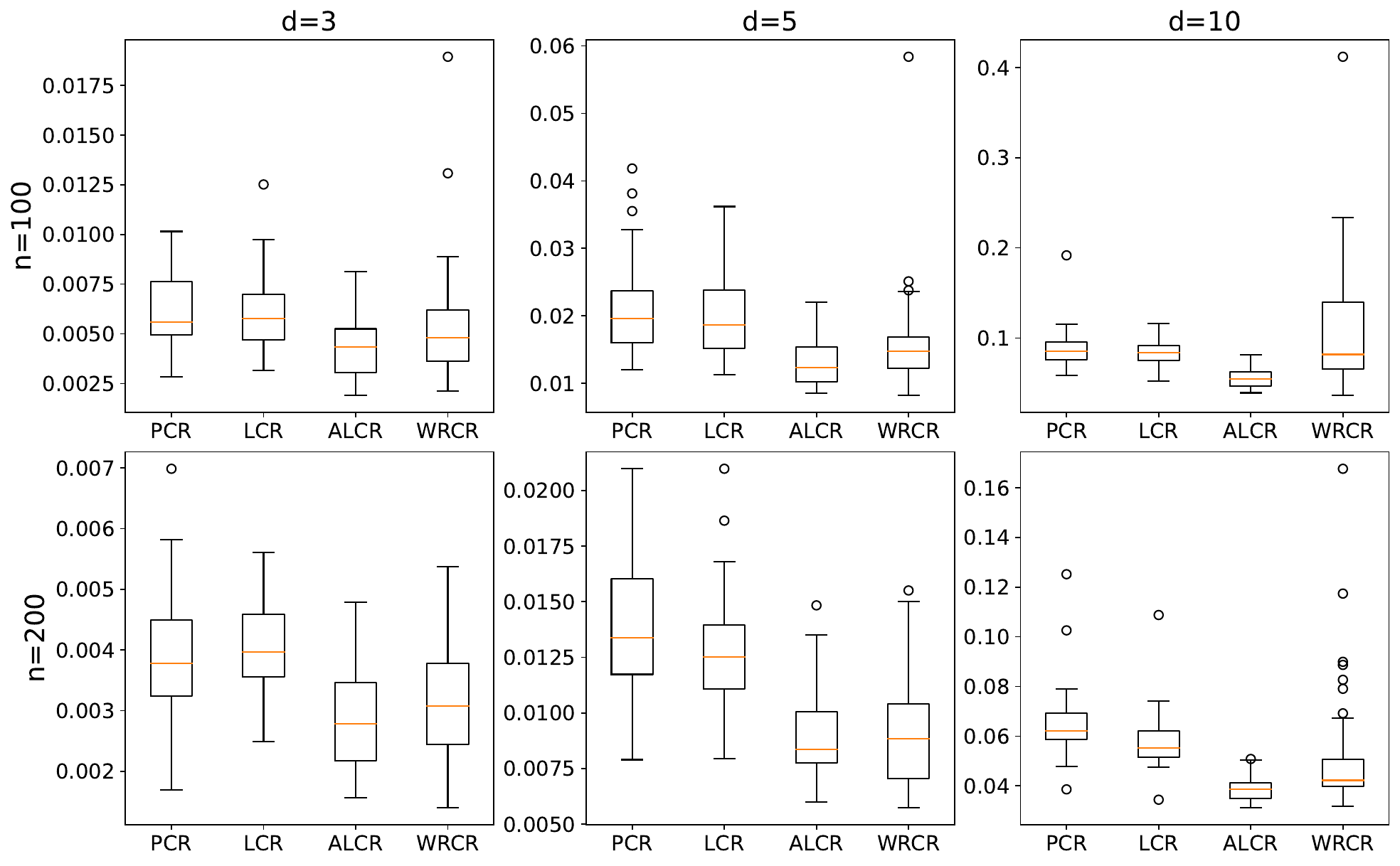}
    \caption{Table showing out-of-sample performance of four methods for data Type II. }
    \label{fig: convb}
\end{figure}

\end{document}